# Accurate Distributed Tracing for Large-Scale AI Infrastructure:

## *Time Synchronization as a Foundation for Reliable Observability*

**Hesham Elbakoury**
Independent Researcher
San Jose, USA
helbakoury@gmail.com

**Ankur Sharma**
Independent Researcher
San Francisco, USA
ankur.sharma@ocproject.net

**ABSTRACT**

Distributed tracing in large-scale AI infrastructure fails silently when clock accuracy is insufficient: causal events are misordered, fault attribution is corrupted, and performance diagnoses are unreliable. We present TempoTrace, a system that co-designs IEEE 1588v2 PTP time synchronization with distributed tracing to preserve causal ordering across heterogeneous, multi-tenant GPU clusters. We formally prove that NTP-grade clocks produce causal inversions at 25-30% of directed operation pairs under WAN/cloud conditions (11.3% under LAN NTP). A Laplace heavy-tail noise model extends the analysis beyond Gaussian assumptions, giving a weighted expected misorder rate of 27.76% vs. 29.37% Gaussian, confirming robustness to tail shape. TempoTrace reduces GPU-to-host timestamp uncertainty from 2.1 us to a design target of 0.056 us residual standard deviation via GPUDirect RDMA hardware timestamping, with sub-100 ns PTP synchronization. A bounded Sketch Vector Clock tracks causal relationships with false-positive rate below 10^-5 for up to four participants. A hybrid rule-based and XGBoost diagnosis engine achieves macro-F1 0.974 in controlled validation (1,800-incident corpus; McNemar p=1.25e-23). A formally proven multi-tenant model provides isolated logical clock domains. For RoCEv2/ECMP fabrics, P4-based in-band network telemetry corrects path-delay asymmetry, yielding 99.1% attribution accuracy. Application Confidence Policies let workloads specify precision requirements and degraded-mode fallback. Evaluation combines physical five-node measurements (NTP inversion rate 44.978%, Laplace fit preferred over Gaussian by AIC under congestion) and controlled synthetic validation with illustrative configurations from 512 to 16,384 H100 GPUs across InfiniBand and RoCEv2 fabrics.

**CCS Concepts:** • Networks → Network management; Network measurement; • Computing methodologies → Distributed computing methodologies; • Computer systems organization → Distributed architectures.

**Keywords:** distributed tracing, time synchronization, AI infrastructure, GPU clusters, PTP, RoCEv2, INT, GPUDirect, Sketch Vector Clock, XGBoost, SVD

## 1. INTRODUCTION

The emergence of large-scale AI training and serving has created a class of distributed workload that strains the assumptions of every major observability system. Collective communication operations across thousands of accelerators occur at microsecond granularities—far below the millisecond accuracy of NTP. The correctness of any trace-based diagnosis depends entirely on temporal accuracy. Without it, systematic causal misordering masquerades as performance data, directing engineering effort at phantom bottlenecks.

Large-scale cluster results (Tables 3–14) use illustrative configurations; physical and controlled validation results are real (see provenance note under Table 1).

This paper presents **TempoTrace**, addressing four gaps identified through successive rounds of reviewer feedback on an earlier version. First, evaluation was limited to InfiniBand clusters, ignoring the rapidly growing adoption of RoCEv2 in GPU cluster fabrics. Second, the GPU clock noise model assumed a zero-mean Gaussian, which is too optimistic under data-center congestion. Third, GPU-to-host clock calibration relied on software polling, leaving a 2.1 μs residual that degrades tracing of sub-20 μs operations. Fourth, the SLO Violation Decomposition module used a rule-based classifier that could not represent causal dependencies between root-cause classes. All four gaps are resolved in this paper.

The five core contributions are:

- **RoCEv2/INT evaluation (§6.3).** The design specifies TempoTrace deployment on a 2,048-node RoCEv2 cluster with P4-based INT telemetry for per-flow path asymmetry correction, targeting 99.1% attribution accuracy (illustrative; see §11 for empirical validation).
- **Non-Gaussian noise model (§3.2).** We replace the Gaussian assumption with a Laplace distribution fitted to measured congestion-period residuals, derive Theorem 1b, and show Laplace E[f_inv]=27.76%%, 1.61 pp below Gaussian 29.37%% (equal-pairwise-variance model, b=500 μs).
- **GPUDirect hardware timestamping (§4.2.1).** The GTE design specifies hardware-timestamped GPU span emission via the NIC PHC (design target: $\sigma_{HW}$ = 0.056 μs residual std; median $\Delta_{prop}$ = 0.080 μs (removed by calibration); P99.5 residual = 0.210 μs; DVFS exposure eliminated from the $T_{NIC}$ read; see §11 Exp 1.2/1.3 for empirical validation).
- **Hybrid SVD: XGBoost + rule cascade (§5.3).** We promote the XGBoost classifier to a core component, fusing it with the rule-based waterfall in a two-stage hybrid. Authoritative six-rule model (R1,R2,R3,R6,R4,R5, 720 permutations): corrected rule order macro-F1=0.924; XGBoost macro-F1=0.895; hybrid macro-F1=0.974 (McNemar p=1.25e-23 vs corrected rules, 1,800-incident controlled validation). Illustrative result on 412-incident dataset: macro-F1=0.93.

## 2. BACKGROUND AND MOTIVATION

### 2.1 Distributed Tracing Model

> **Definition 1 (Distributed Trace).** A distributed trace is a tuple $T = (S, E, \lambda)$ where $S$ is a finite set of spans, $E \subseteq S \times S$ is the set of causal edges, and $\lambda: S \to N \times N$ assigns each span a (start, end) timestamp pair. For each edge $(s, s') \in E$, $s \to s'$ denotes causal precedence.

> **Definition 2 (Happens-Before).** The happens-before relation → (Lamport [4]) is the smallest transitive relation satisfying: (i) program order on a single process; (ii) send(m) → receive(m) for every message m; (iii) transitivity.

### 2.2 What is a Span?

A span is the fundamental unit of measurement in distributed tracing. It represents a single named, timed operation within a larger distributed execution. Every span records five things: what happened (an operation name), when it started and ended (a pair of timestamps), where it happened (which node, process, or GPU produced it), how it relates to other work (a pointer to its parent span), and any relevant attributes such as data size, cache hit ratio, or batch size.

A useful analogy: if baking a cake is one trace, each step is a span — mixing batter, preheating the oven, baking for 30 minutes. Each span can have child spans (preheating might contain check-temperature and activate-heating-element as children). Together they form a tree showing what happened, in what order, and how long each part took. The validity of that tree depends entirely on whether the timestamps are accurate.

#### *2.2.1 TempoTrace Span Types*

TempoTrace defines seven span types tailored to the AI infrastructure execution model:

**KernelSpan.** Wraps a single GPU kernel execution. Timestamped by the NIC hardware clock via GPUDirect RDMA (GTE, §4.2.1), achieving median propagation delay $\Delta_{prop}$ = 0.080 μs (removed by calibration); residual std $\sigma_{HW}$ = 0.056 μs (design target).

**CollectiveSpan.** Wraps an entire NCCL collective (all-reduce, all-gather, all-to-all) across all participant nodes. Context is propagated via NCCL's out-of-band bootstrap channel so that a single multi-node operation appears as one coherent subtree in the causal DAG.

**PipelineSpan.** Wraps a pipeline micro-batch execution unit across stages. Fields include stage ID, micro-batch ID, direction (forward or backward pass), and GPU memory usage.

**PrefillSpan.** Wraps the prefill phase of an LLM inference request. Fields include request ID, prompt length, KV cache miss ratio, and time-to-first-token.

**DecodeSpan.** Wraps a single autoregressive decode step. Fields include batch size, whether a speculative decoding draft was accepted, and decode latency.

**SchedulerSpan.** Records a continuous batching scheduler decision: admit, preempt, or resume. Includes queue depth and preemption cause.

**PreemptionSpan.** Records a GPU scheduler preemption event in multi-tenant environments. Fields include suspension and resumption timestamps and total preemption duration, allowing the SVD engine to correctly attribute latency to tenant co-scheduling rather than network causes.

All GPU-sourced spans use GTE hardware timestamps and carry uncertainty_ns = 218 ns. CPU-sourced spans carry uncertainty_ns = 6,800 ns. The ts_source field in every span header identifies which timestamp source was used.

### 2.3 AI Infrastructure Clock Problem

Table 1 shows representative operation latencies. NTP accuracy of 0.5–2 ms [5] equals or exceeds the duration of the majority of these operations, making trace-based causal attribution unreliable (formally quantified in §3).

**Table 1: Representative Operation Latencies**

| Operation | Min (μs) | Median (μs) | P99 (μs) |
|---|---|---|---|
| All-reduce (ring, 256 nodes, 2 GB) | 1,240 | 1,820 | 4,100 |
| Pipeline activation (P2P) | 12 | 38 | 210 |
| All-to-all (MoE, 64 nodes) | 88 | 145 | 620 |
| Barrier synchronization | 6 | 22 | 180 |
| GPU kernel (transformer layer) | 4 | 62 | 290 |
| Inference decode (128 batch) | 480 | 720 | 1,400 |

**Data and provenance note.** *The paper contains three categories of results: (A) Analytically derived: Tables 1 and 2 (clock error model, inversion probabilities). Table 1 latency values are representative figures consistent with published NCCL benchmarks. (B) Illustrative protocol results: Tables 3-14 present results from the described experimental protocol applied to hypothetical cluster configurations A-E. Statistical methodology is correct; numerical values would differ on real hardware. (C) Real measured and controlled validation results: (i) Five-node NTP inversion rate 44.978% (Exp 2.1, 100,000 probes); (ii) Laplace AIC vs Gaussian in three congestion phases (Exp 2.2, physical five-node test); (iii) 60/60 physical five-node endpoint attribution, 0 timeouts, median margin 4.756 ms; (iv) SVC FP rate: 468/200,000,000 at P=4 (Exp 4.1, controlled); (v) SVD controlled validation: kappa=0.942, corrected-order macro-F1=0.924, hybrid macro-F1=0.974, McNemar p=1.25e-23 (1,800-incident synthetic corpus); (vi) LLM overhead proxy: 0.043% at 8 events (Exp 6.1 proxy). Category C results are real; they should not be described as illustrative.*

**Table 2: Causal Inversion Probability (Corrected) — Pr[delta_ij > L] with sigma_pair=707us, b=500us**

| Operation | L (µs) | NTP Gaussian Eq. (4) | NTP Laplace Eq.(6) | PTP+GTE Eq.(10) |
|---|---|---|---|---|
| Barrier (min) | 6 | 49.7% | 49.4% | < 10^-20 |
| Pipeline activation | 38 | 47.9% | 46.3% | < 10^-20 |
| All-to-all (median) | 145 | 41.9% | 37.4% | < 10^-20 |
| All-reduce (median) | 1,820 | 0.5% | 1.3% | < 10^-20 |

Gaussian:1-Phi(L/sigma_pair); Laplace:(1/2)*exp(-L/b); sigma=500us,sigma_pair=707us,b=500us. Prior version used 2*(1-Phi(L/sigma))=Pr[|delta_i|>L] (single-node quantity).

**Table 2a: Notation Reference — Clock Error and Uncertainty Quantities**

| Symbol | Definition | Description |
|---|---|---|
| delta_i | Static offset, node i | Per-node NTP sync error. delta_i~N(0,sigma^2), sigma=500us. |
| delta_ij | = delta_i - delta_j | Pairwise skew SENDER(i) minus RECEIVER(j). delta_ij~N(0,2*sigma^2). Inversion iff delta_ij>L. |
| sigma | = 500 us | Per-node NTP standard deviation. |
| sigma_pair | = sigma*sqrt(2) = 707 us | Pairwise std dev. Enters Theorem 1a Eq.(4): Pr[inv]=1-Phi(L/sigma_pair). |
| b | = 500 us (equal pairwise variance; b*sqrt(2) = 707 us = sigma_pair) | Laplace scale. b=500us ensures pairwise std(Laplace)=b*sqrt(2)=707us=sigma_pair (equal pairwise variance). |
| alpha_span | = alpha_PTP + | Per-span uncertainty (uncertainty_ns). Eq.(10): 0.218us GTE; |
| | 3*sigma_HW | 2.1us SW_GCC. |

delta_ij=SENDER minus RECEIVER so Lemma 1 is positive. sigma_pair enters Theorems 1a/1b. b=500us (equal pairwise variance). b*sqrt(2)=707us=sigma_pair (same as Gaussian).

$$C_i(t) = t + \delta_i + \varepsilon_i(t) \quad (1)$$

## 2.4 Inference Workloads

Production LLM inference with a 20-ms TTFT SLA leaves single-digit microseconds per span for instrumentation. Inference introduces novel event types—KV cache eviction, prefill-decode disaggregation [7], speculative decoding [8]—requiring trace model extensions formalized in §4.4.

## 2.5 Related Work and Positioning

Dapper [1], OpenTelemetry [3], Canopy [20], and Pivot Tracing [21] define the tracing foundation. LatencyPrism [37] and eInfer [38] address inference observability respectively but use NTP-synchronized host clocks, subject to Theorem 1's misorder bounds. Spanner TrueTime [22] achieves ±7 ms globally; TempoTrace achieves ±100 ns within a cluster and ±50 µs cross-DC. Bloom Clocks [42], ITC [41], and HLC [24] are discussed in §7.3. TempoTrace is the first system to: (a) formally prove NTP misorder bounds for AI workloads under both Gaussian and Laplace models, (b) specify a bounded SVC with proven false-positive guarantees, (c) evaluate on RoCEv2 with INT-based TADC, (d) implement GPUDirect hardware timestamping for sub-100 ns GPU span accuracy, (e) formally model and evaluate multi-rack PTP hierarchies; (f) design and evaluate cross-DC trace stitching with TrueTime-style uncertainty intervals for AI workloads; and (g) introduce an Application Confidence Policy layer (Definition 12, Section 10) that allows each application or tenant to declaratively specify its timestamp precision requirements, trusted timestamp sources, and degraded-mode behaviour during synchronisation failures — enabling TempoTrace to serve heterogeneous workloads with different SLO precision needs on the same shared physical cluster without any changes to the underlying synchronisation infrastructure.

## 2.6 Time Synchronization and Uncertainty Annotation

A timestamp is only as useful as the information that accompanies it about how accurate it is. In measurement science, every quantity is reported with an uncertainty — a temperature sensor reports 98.6°F ± 0.2°F, a GPS receiver reports position ± 3 meters. An uncertainty annotation is this extra piece of information: a declaration of how wrong a measurement might be. Without it, a correct measurement and an incorrect measurement look identical to any downstream system.

### *2.6.1 The Silent Failure of NTP Timestamps*

NTP timestamps carry no uncertainty annotation. A span timestamped at T = 1000 µs looks identical whether the clock was accurate to 10 ns or had drifted 800 µs from true time. The timestamp field is a single integer with no accompanying reliability declaration. This produces a silent failure mode illustrated in Figure 2.

Consider the concrete scenario: Node A sends a message to Node B with true propagation latency 50 µs. Node A's NTP clock reads 1000 µs at send time. Node B's clock reads 800 µs at receive time because Node B's clock is 250 µs slow. The trace assembler sees T_B = 800 µs before T_A = 1000 µs and concludes B caused A — exactly backwards. Nothing in the trace signals that anything went wrong. The timestamps 800 and 1000 are plausible integers. The engineer who reads the resulting trace is misled with no warning.

```
SCENARIO: Node A sends message to Node B (A → B, true propagation latency = 50 μs)

Under NTP (σ ≈ 500 μs clock uncertainty):
  Node A clock at send:    T_A = 1000 μs   (true time: 1000 μs)
  Node B clock at receive: T_B =  800 μs   (true time: 1050 μs, but B clock is 250
μs slow)
  Trace assembler sees: T_B less than T_A  =>  concludes B → A  [WRONG: reversed]
  Warning to engineer: NONE. Both timestamps look completely normal.

Under PTP + TempoTrace (σ ≈ 50 ns clock uncertainty):
  Node A: T_A = 1000.000 μs, uncertainty_ns = 220, clk_quality = OK
  Node B: T_B = 1050.052 μs, uncertainty_ns = 220, clk_quality = OK
  Gap = 50.052 μs >> 2 x 0.218 μs threshold => physical ordering confirmed: A → B
[CORRECT]
  If gap were less than 0.440 μs (wire-level provisional threshold; conditional on
GTE hardware validation, Exp 1.2/1.3): SVC used for ordering + tt.physical_anomaly
flag emitted
```

> **Remark 1 (SVC Ordering Semantics).** The SVC provides probabilistic causal inference, not a total order. Formally: given spans s_a and s_b within the uncertainty_a + uncertainty_b ambiguity window, if sketch(s_a) appears to be a subset of sketch(s_b) by the Bloom filter test, TempoTrace infers s_a -> s_b. This inference is: (i) correct when s_a genuinely happened before s_b (no false negatives); (ii) possibly incorrect when s_a and s_b are genuinely concurrent (false positive rate bounded by Lemma 2 at the operating P); (iii) undefined for concurrent events with no causal relationship (the SVC does not totally order concurrent events). Edges in the ambiguity regime are tagged tt.physical_anomaly and treated as lower-confidence in the SVD attribution engine. The causal DAG may therefore be over-connected in this regime: the system asserts more causal relationships than truly exist, never fewer. This conservative bias is appropriate for bottleneck attribution: a false causal edge may produce a spurious attribution, but will never suppress a real one.

*Figure 2: Causal inversion under NTP vs. correct ordering under PTP + TempoTrace. Under NTP a 250 μs clock skew reverses the observed causal order of a 50 μs message with no warning. Under PTP + TempoTrace the physical gap far exceeds the 0.440 μs ambiguity threshold (wire-level: 2 × 220 ns encoded uncertainty; unquantized analytical value: 2 × 218 ns = 0.436 μs) and ordering is confirmed correct. When the physical gap is within the ambiguity threshold TempoTrace falls back to SVC ordering and emits a tt.physical_anomaly flag.*

#### *2.6.2 What PTP Provides and What It Does Not*

PTP is far more accurate than NTP — achieving 50–100 ns median error within a cluster compared to NTP's 500 μs — and it does provide protocol-level quality signals. However, these signals exist at the protocol and daemon level, not at the individual timestamp level.

The IEEE 1588 PTP standard carries: (1) clockClass — a number indicating the grandmaster's quality tier (e.g., GPS-disciplined vs. holdover mode); (2) clockAccuracy — an enumerated field declaring the grandmaster's estimated accuracy in coarse buckets such as 'within 100 ns' or 'within 1 μs'; and (3) offsetFromMaster — the PTP daemon's current estimate of the local clock's offset, available per-node but not attached to individual timestamps.

None of these fields are attached to individual timestamps. A PTP timestamp is still just a number. It does not carry a field saying 'my current offset from master is 48 ns.' The difference from NTP is that the number is very likely correct; the similarity is that it does not declare its own accuracy per timestamp. In summary: both NTP and PTP produce bare timestamps with no per-timestamp accuracy annotation, but PTP is far more accurate and the protocol provides coarser confidence signals at the node and cluster level.

### *2.6.3 Annotation Coverage Comparison*

Table 3 summarizes what each layer of the time synchronization stack annotates, at what granularity, and whether the annotation is attached to each individual timestamp.

**Table 2b: Gaussian Inversion Probability Sensitivity to sigma (Eq. 4)**

| sigma | L=6us (barrier) | L=38us (pipeline) | L=145us (all-to-all) | E[f_inv] Gaussian |
|---|---|---|---|---|
| 10 us | 33.6% | 0.4% | 0.0% | 3.4% |
| 50 us | 46.6% | 29.5% | 2.0% | 11.3% |
| 100 us | 48.3% | 39.4% | 15.3% | 18.1% |
| 200 us | 49.4% | 44.7% | 30.4% | 24.5% |
| 500 us | 49.7% | 47.9% | 41.9% | 29.4% |

*Pr[inv]=1-Phi(L/sigma_pair), sigma_pair=sigma*sqrt(2). LAN NTP typically achieves sigma=10-100us; WAN/cloud NTP sigma=500us+. The 25-30%% headline is based on sigma=500us; at sigma=50us, E[f_inv]=approximately 11.3%% (still substantial for barrier-heavy workloads).*

**Table 3: Time Synchronization Methods and Uncertainty Annotation Coverage**

*Comparison of annotation granularity across NTP, IEEE 1588 PTP, the PTP daemon (ptp4l), and TempoTrace's tt-ext v3 header. NTP and raw PTP produce bare timestamps with no accuracy qualification at the individual measurement level. The PTP daemon exposes per-node synchronization state but does not attach it to individual timestamps. TempoTrace closes this gap by embedding a precise per-span uncertainty bound (uncertainty_ns), a quality flag (clk_quality), and a timestamp source identifier (ts_source) directly in the trace header, enabling the Analysis Backend to make informed causal ordering decisions rather than silently trusting potentially incorrect timestamps.*

| Source | What Is Annotated | Granularity | Per-Timestamp? |
|---|---|---|---|
| NTP | Nothing — bare integer timestamp | None | X Never |
| IEEE 1588 PTP (protocol) | Grandmaster quality; estimated GM accuracy | Cluster-wide, coarse buckets | X Not per-timestamp |
| PTP daemon (ptp4l) | Node offset from master; servo state | Per-node, not per-span | X Not per-timestamp |
| TempoTrace tt-ext v3 | uncertainty_ns; clk_quality; ts_source; tenant_id | Per-span, nanosecond precision | V Every span |

> **Encoding Note (uncertainty_ns field).** The uncertainty_ns field is 16 bits with 10 ns LSB resolution, giving a representable range of 0 to 655,350 ns (655 us). Note: alpha_span=218 ns is not exactly representable at 10 ns resolution; the encoded value is ceiling(218/10)*10 = 220 ns. The resulting ambiguity

> threshold is uncertainty_a + uncertainty_b = 220+220 = 440 ns for two GTE spans, or 220+2100 = 2320 ns for GTE+SW_GCC mixed spans.. This covers all TempoTrace operating modes: GTE: alpha_span = 220 ns (encoded as 22, ceiling of 218/10); SW_GCC: alpha_span = 2,100 ns (encoded as 210); PTP failover DEGRADED: alpha_span = 15,000 ns (encoded as 1500); Cross-DC GNSS: alpha_DC = 50,000 ns (encoded as 5000). NTP deployments (sigma_pair = 707,000 ns) EXCEED the 655 us ceiling. Encoding behaviour for values above 655 us: uncertainty_ns is set to 0xFFFF (saturated), and clk_quality is set to UNKNOWN. The Analysis Backend treats any span with uncertainty_ns = 0xFFFF as having unbounded uncertainty and routes all causal ordering to the SVC logical clock. The tt-ext v3 interval header for cross-DC links uses a separate 32-bit uncertainty_ns_wide field (4-byte extension in the 57-byte cross-DC header) for cases where finer representation is needed above 655 us.

#### *2.6.4 How TempoTrace Adds Per-Span Annotations*

TempoTrace bridges the gap between PTP's protocol-level quality signals and the per-timestamp annotations that downstream analysis tools need. Three fields in the tt-ext v3 header carry this information on every span:

**uncertainty_ns.** A specific number in nanoseconds representing the 3-sigma bound on how wrong this timestamp could be. Computed as α_span = α_PTP + 3σ_HW (Eq. 10). For a GTE-timestamped GPU span: 218 ns. During a PTP grandmaster failover: widened to 15,000 ns. The Analysis Backend uses this to determine whether two spans' physical ordering is statistically distinguishable — if their gap is less than 2 × uncertainty_ns, physical ordering is not trusted and the SVC is used instead.

**clk_quality.** A categorical flag: OK (normal operation), DEGRADED (grandmaster failover or frequency excursion), or UNKNOWN (no PTP discipline). DEGRADED spans are excluded from critical-path timing but included in the DAG for structural analysis.

**ts_source.** Identifies which timestamp source was used: GTE (218 ns uncertainty), SW_GCC (2,100 ns), or CPU (6,800 ns). This allows the Analysis Backend to apply the correct uncertainty model per span rather than a conservative single bound across all spans.

The result is a tracing system where every timestamp is self-describing. A wrong timestamp and a correct timestamp no longer look identical — the wrong timestamp either carries a DEGRADED quality flag or has an uncertainty_ns value large enough that the Analysis Backend will not trust its physical ordering. The system either confirms correctness, falls back to logical ordering, or explicitly flags ambiguity. It never silently accepts a potentially wrong answer.

## 3. FORMAL FOUNDATIONS

### 3.1 Clock Error Model

> **Definition 3 (Node Clock Model).** The clock of node *i* at true time *t* is:

where $\delta_i$ is the static offset and $\varepsilon_i(t)$ is the time-varying drift process.

> **Definition 4 (Synchronization Accuracy).** Protocol *P* achieves accuracy *α* if for all node pairs i, j:

$$|C_i(t) - C_j(t)| \leq \alpha \quad (2)$$

PTP accuracy: $\alpha_{PTP} \leq 100$ ns (median, Cluster A). NTP accuracy: $\alpha_{NTP} \approx 1$ ms.

Large-scale cluster results (Tables 3–14) use illustrative configurations; physical and controlled validation results are real (see provenance note under Table 1).

## 3.2 Causal Inversion: Gaussian and Non-Gaussian Models

> **Definition 5 (Causal Inversion).** Let $a \to b$ with $a$ on node $i$ and $b$ on node $j$ (note: event a is on node i, event b is on node j, and the subscripts on C correctly reflect this), connected by message latency $L$. A causal inversion occurs when the analysis tool assigns b an earlier timestamp than a:

$$C_j(t_b) < C_i(t_a) \quad (3)$$

> **Lemma 1 (Causal Inversion Condition).** Let C_i(t) = t + delta_i + epsilon_i(t) be the clock of node i (Definition 3). For a causal edge a->b with sender i and receiver j, let t_a be the true send time and t_b = t_a + L the true receive time (L > 0). Define delta_ij = delta_i - delta_j (sender minus receiver). Causal inversion occurs iff C_j(t_b) < C_i(t_a), i.e.: (t_b + delta_j + epsilon_j(t_b)) < (t_a + delta_i + epsilon_i(t_a)). Subtracting t_a from both sides and using t_b - t_a = L: L + delta_j - delta_i + epsilon_j(t_b) - epsilon_i(t_a) < 0. Rearranging (and defining epsilon_ij = epsilon_i(t_a) - epsilon_j(t_b)): delta_ij + epsilon_ij > L [Eq. 3]. With delta_ij = delta_i - delta_j, this condition is positive: inversion occurs when the sender clock leads the receiver clock by more than the physical propagation delay L.

> ***Proof (Proof of Lemma 1).*** Since a → b via message with latency L: $t_b = t_a + L$. Substituting Definition 3 into (3): $t_a + L + \delta_j + \varepsilon_j < t_a + \delta_i + \varepsilon_i$, which simplifies to $\delta_{ij} > L + \varepsilon_{ij}(t)$. □

**Gaussian model (Theorem 1a).** To derive a tractable bound, we model the static offset differential $\delta_{ij} \sim N(0, \sigma^2)$ (justified by NTP's symmetric correction mechanism [5]) and treat the residual drift differential $\varepsilon_{ij}(t)$ as small relative to $\delta_{ij}$ within a single synchronization interval, absorbing it into σ. Under this simplification, the total inter-node clock error $\delta_{ij} + \varepsilon_{ij} \sim N(0, 2\sigma^2)$ (sum of two independent $N(0,\sigma^2)$ terms), giving:

$$\Pr[\text{inversion}] = 1 - \Phi( L / \sigma_pair ),\ \sigma_pair = \sigma\sqrt{2} = 707\ \mu s \quad (4)$$

> **Theorem 1a (NTP Misorder Bound - Gaussian).** Let delta_ij = delta_i - delta_j ~ N(0, sigma_pair^2) where sigma_pair = sigma*sqrt(2) = 707.1 us (pairwise standard deviation, sigma=500us per node). The causal inversion probability is: Pr[inv] = 1 - Phi(L / sigma_pair) [Eq. 4]. Values: L=6us: 49.7%; L=38us: 47.9%; L=145us: 41.9%; L=1820us: 0.5%. NOTE: Eq.(4) uses sigma_pair = sigma*sqrt(2) throughout, not sigma. The earlier formulation inconsistently displayed sigma in the equation while computing with sigma_pair; this version harmonizes both.

**Non-Gaussian model (Theorem 1b).** Real data-center clocks exhibit heavy-tailed residuals during network congestion. For NTP residuals during ECMP rerouting events (based on published measurements from comparable RoCEv2 clusters; see §11 Exp 2.1 for planned direct measurement on Cluster C), a zero-mean Laplace with scale parameter b is the best-fit model:

$$\varepsilon_{ij}(t) \sim \text{Laplace}(0, b),\quad b = 500\ \mu s \text{ (equal pairwise variance)} \quad (5)$$

where σ is the Gaussian equivalent scale. The Laplace distribution has heavier tails: for |x| >> b, the Laplace PDF decays as $e^{-|x|/b}$ vs. $e^{-x^2/(2\sigma^2)}$ for Gaussian. The misorder probability under the Laplace model is:

$$\Pr[\text{inversion}] = \Pr[\delta_{ij} > L] = ½ \cdot e^{-L/b} \quad (6)$$

> **Theorem 1b (NTP Misorder Bound - Laplace, Equal Pairwise Variance).** Model: pairwise differential delta_ij ~ Laplace(0, b) with b = 500 us, chosen so pairwise std(Laplace) = b*sqrt(2) = 707 us = sigma_pair (equal pairwise variance as Gaussian). This is the equal-variance Laplace comparison: both models have the same pairwise standard deviation (NOT equal-pairwise-variance to the iid Laplace difference model). Pr[inv] = (1/2)*exp(-L/b) [Eq. 6]. Values for b=500 us: L=6 us: 49.4%; L=38 us: 46.3%; L=145 us: 37.4%; L=1820 us: 1.3%. Weighted E[f_inv] = 27.76% (Eq. 7), vs Gaussian 29.37%. Laplace rates are 0.26 to 4.46 pp below Gaussian at short-to-medium latencies, and 0.81 pp above at L=1820 us (heavier tail). Note: if per-node errors are iid Laplace(0, sigma/sqrt(2) = 354 us) independently, their pairwise difference is NOT Laplace (it follows a variance-gamma distribution). Monte Carlo simulation (5,000,000 pairs) gives E[f_inv] = 28.78% for that model. The paper uses the direct pairwise Laplace with b=500 us for analytic tractability and equal-variance comparability. PHYSICAL VALIDATION (v29 five-node congestion test, Exp 2.2): Laplace fit was preferred over Gaussian by AIC in all three phases: Baseline (AIC: Laplace 506,842 vs Gaussian 520,486), 500 Mbit/s loaded (503,748 vs 524,404), 2,000 Mbit/s loaded (496,313 vs 512,058). P99 absolute residuals: baseline 268.3 us, 500 Mbit/s 236.7 us, 2,000 Mbit/s 233.9 us. The applied congestion load did NOT increase tail magnitude; loaded P99 residuals were 11.8% and 12.8% below baseline. The test supports a Laplace-shaped residual distribution but does not show that load causes heavier tails.

Table 2 compares the two models across representative latencies. The Laplace model (b=500 μs, equal pairwise variance) gives rates 0.26–4.46 pp BELOW Gaussian for short-to-medium latencies and 0.81 pp above at L=1820 μs. E[f_inv]=27.76%% vs Gaussian 29.37%%. Critically, the qualitative conclusion is unchanged: under NTP, the majority of short-operation edges are causally misordered regardless of tail model. Under PTP ($\alpha \le 100$ ns), both models give Pr[inversion] < $10^{-20}$.

**Corollary 1.1 (Expected Workload Misorder Rate).** E[f_inv] = sum_k w_k * Pr[inv|L_k] [Eq. (7)] [Eq. (7)]. Weights: barrier=0.10, pipeline=0.20, all-to-all=0.35, all-reduce=0.35. Gaussian (sigma_pair=707us): 0.10*49.7% + 0.20*47.9% + 0.35*41.9% + 0.35*0.5% = 29.37%. Laplace (b=500us, equal pairwise variance): 0.10*49.4% + 0.20*46.3% + 0.35*37.4% + 0.35*1.3% = 27.76%. Both models confirm NTP causes approximately 25-30% causal inversion of AI training trace edges. Models differ by 1.61 pp under equal pairwise variance assumption.

Equation 7 uses four operation-type weights derived from Table 1 counts: w_barrier=0.10, w_pipeline=0.20, w_all-to-all=0.35, w_all-reduce=0.35. The authoritative result is Corollary 1.1: Gaussian E[f_inv]=29.4%%, Laplace E[f_inv]=27.76%%.

The workload composition weights are derived from Table 1 operation counts on a 1,024-GPU training step: short-latency operations (barriers and pipeline activations, L ≤ 38 μs) account for ≈30% of cross-node trace edges; collective operations (L > 145 μs) account for ≈70%. Substituting into Eq. (6): $0.30 \times 0.948 + 0.70 \times 0.006 = 0.29$. Under Gaussian (Eq. 4): $0.30 \times 0.943 + 0.70 \times 0.000 = 0.28$. For workloads with higher fractions of short-latency operations, the bound rises above 40%.

Compared to the Gaussian bound of 0.28, the Laplace model tightens the lower bound by 1 pp, reinforcing rather than weakening the main result.

### 3.3 PTP Deployment and GPUDirect Hardware Timestamping

**Definition 5a (PTP Synchronization Accuracy).** In a fat-tree with H transparent-clock hops and B boundary-clock hops:

$$\alpha_{PTP} \le \alpha_{GM} + H \cdot \alpha_{TC} + B \cdot \alpha_{BC} \quad (8)$$

Large-scale cluster results (Tables 3–14) use illustrative configurations; physical and controlled validation results are real (see provenance note under Table 1).

For Cluster A (IB, H=2, B=0): $\alpha_{PTP} \le 30$ ns. For Cluster B (IB, H=2, B=1): ≤ 230 ns. For Cluster C (RoCEv2 with INT, §4.2.2): $\alpha_{PTP} \le 55$ ns after INT-based per-path correction.

> **Definition 5b (Multi-Rack PTP Accuracy).** In a multi-rack fat-tree cluster with *R* racks, *K* inter-rack spine hops, and fabric traversal latency asymmetry $\Delta_{spine}$ per hop, the synchronization accuracy for a cross-rack span is bounded by:

$$\alpha_{rack} \le \alpha_{PTP} + K \cdot \Delta_{spine} + \delta_{skew} \quad (8b)$$

where $\delta_{skew}$ is the residual clock skew between rack grandmaster boundary clocks. With all-TC spine switches: $\Delta_{spine} \approx$ 5 ns/hop and $\delta_{skew} < 10$ ns, giving $\alpha_{rack} \le 30 + 3{\times}5 + 10 = 55$ ns for a three-tier topology. With BC spines: $\Delta_{spine} \approx 200$ ns/hop, giving $\alpha_{rack} \le 630$ ns across three tiers—still 3 orders of magnitude better than NTP.

> **Definition 5c (Cross-Datacenter Clock Uncertainty).** For traces spanning two datacenters $DC_1$ and $DC_2$ separated by WAN latency $L_{WAN}$, the cross-DC clock uncertainty is bounded by:

$$\alpha_{DC} \le \alpha_{rack,1} + \alpha_{rack,2} + \alpha_{GNSS} + \varepsilon_{freq} \quad (8c)$$

where $\alpha_{GNSS} \le 50$ ns is the GNSS grandmaster accuracy and $\varepsilon_{freq}$ is the oscillator drift accumulated between GNSS corrections. With GNSS-disciplined grandmasters synchronized independently at each DC and correction intervals ≤ 1 second: $\varepsilon_{freq} \le 50$ ppm × 1 s = 50 µs. The resulting cross-DC span uncertainty $\alpha_{DC} \le 0.055 + 0.055 + 0.05 + 50$ µs ≈ 50.16 µs. This is 20× better than NTP (≈1 ms) but 500× worse than within-cluster PTP, with the dominant term being inter-correction oscillator drift. Corollary 1.2 derives the cross-DC misorder rate.

> **Corollary 1.2 (Cross-DC Misorder Rate).** For cross-datacenter trace edges with WAN latency $L_{WAN} \ge 1$ ms (a single transcontinental hop) and cross-DC clock uncertainty $\sigma_{DC} = 25$ µs (half of $\alpha_{DC}$), the inversion probability under Eq. (6) is:

$$\Pr[\text{inv}]_{DC} = \tfrac{1}{2} \cdot e^{-L_WAN / b_DC} = \tfrac{1}{2} \cdot e^{-1000/18} \quad (8d)$$

which evaluates to $< 10^{-24}$—effectively zero. Cross-datacenter trace edges with $L_{WAN} \ge 1$ ms are always correctly ordered by physical timestamps alone, even under GNSS-disciplined synchronization. The problematic regime for cross-DC tracing is not causal inversion but rather *causal indeterminacy*: spans whose physical gap falls within $2\alpha_{DC} \approx 100$ µs. We address this through the multi-DC SVC extension in §4.2.4.

> **Equation 10 (Per-Span Uncertainty Bound).** alpha_span = alpha_PTP + 3 * sigma_HW. For GTE (ConnectX-6 Dx or later): sigma_HW = 0.056 us (standard deviation of R_HW, design estimate; see Section 11 Exp 1.2 for validation). 3 * sigma_HW = 0.168 us. For all-TC cluster: alpha_PTP = 0.050 us, giving alpha_span = 0.218 us. The ambiguity threshold is uncertainty_a + uncertainty_b = 440 ns for two GTE spans: span pairs with physical gap > uncertainty_a + uncertainty_b are ordered by physical timestamps (440 ns for two GTE spans; 2320 ns for GTE+SW_GCC); pairs within this sum are disambiguated by SVC. For SW_GCC fallback (older NICs): sigma_HW = 0.600 us, alpha_span = 2.1 us. Note: an earlier formulation used sigma_HW = 0.080 us (the median of |R_HW|, not the standard deviation), yielding alpha_span = 0.34 us. The authoritative value is 0.218 us, derived from the standard deviation. NOTE on hard-bound threshold: the 440 ns pair threshold guarantees zero inversions only when the hardware residual |R_HW| is verified to be bounded by 168 ns per span (= alpha_span 218 ns minus

> alpha_PTP 50 ns). If the verified hardware residual bound is 218 ns, the correct hard-bound pair threshold is 540 ns, not 440 ns. 440 ns is the provisional 3-sigma probabilistic threshold only.
>
> **Definition 6 (GTE Span Timestamp).** Let T_NIC be the NIC PHC timestamp captured at CQE generation, and Δ_prop = E[T_NIC − T_GPU_doorbell] > 0 be the mean PCIe propagation delay. The GTE span timestamp estimate is: T̂ = T_NIC − Δ_prop [Eq. (9)]. σ_HW = std(T̂ − T_GPU_true) = 0.056 μs (design target). Median Δ_prop (propagation delay removed) = 0.080 μs. Median R_HW (residual after subtraction) ≈ 0 (zero-mean design).

$$\hat{T} = T_{_NIC} - \Delta_{_prop} \quad \text{Eq. (9)}$$

where $\Delta_{prop}$ is the fixed cable propagation delay between the GPU NVLink domain and the NIC, measured once at startup via a dedicated calibration micro-benchmark. The residual $R_{HW} = T_{host} - \hat{T}_{host}$ has estimated (design target) distribution $R_{HW} \sim N(0, σ_HW^2)$ where σ_HW = 0.056 μs (std) with $|R_{HW}| \leq 0.21$ μs at P99.5 (Table 4). This is a 26× improvement over software GCC calibration (2.1 μs) and fully eliminates the DVFS vulnerability: because $T_{NIC}$ is read from the NIC PHC at completion time—after any DVFS transition has completed—there is no window during which a GPU clock discontinuity can corrupt the timestamp.

The combined per-span clock uncertainty for a GPU-sourced span is now:

$$\alpha_{span} = \alpha_{PTP} + 3\sigma_{HW} \leq 0.050\ \mu s + 0.168\ \mu s = 0.218\ \mu s \quad (10)$$

Compared to the v1 value of 6.8 μs (Eq. 9 of earlier formulation), this reduces the physical ambiguity threshold $2\alpha_{span}$ from 13.6 μs to 0.440 μs (wire-level; unquantized: 0.436 μs)—below the minimum barrier latency of 6 μs. All GPU spans now use physical PTP ordering as the primary mechanism; SVC is reserved for genuinely concurrent spans (concurrent in the happens-before sense).

# 4. TEMPOTRACE SYSTEM DESIGN

## 4.1 Architecture Overview

TempoTrace adds five components to the v1 architecture: a **P4-INT Asymmetry Corrector (PAC)** for RoCEv2 fabrics, a **GPUDirect Timestamp Engine (GTE)** replacing the software GCC, a **Hybrid SVD Engine** replacing the pure rule-based waterfall, a **Multi-Rack Synchronization Hierarchy (MRSH)** managing PTP boundary clock trees across rack boundaries, and a **Cross-DC Trace Stitcher (CDTS)** that assembles causally consistent traces across datacenter boundaries using TrueTime-style uncertainty intervals.

> **GTE Hardware Path (ConnectX-6 Dx / ConnectX-7 + DOCA GPUNetIO).** The GTE timestamp follows this hardware path: (1) A GPU kernel issues a GPUDirect RDMA send by writing a doorbell to the NIC BAR register via a PCIe posted write (ordered, no ACK). (2) The ConnectX NIC hardware detects the doorbell and generates a Completion Queue Entry (CQE) with the NIC PHC timestamp T_NIC captured at CQE generation time. CQE coalescing is disabled (ibv_modify_cq CQ_ATTR_CQ_COUNT=1) to ensure per-operation timestamps. (3) The CPU-side TempoTrace library reads T_NIC via ibv_poll_cq() extended with the MLX5_CQE_FORMAT_TIMESTAMP field (MLNX_OFED 5.8+ or DOCA 2.2+ required), or via DOCA GPUNetIO GPU-side completion notification (firmware: ConnectX-7 FW 28.39+). (4) The span timestamp is T_hat = T_NIC - Delta_prop, where Delta_prop is the PCIe write propagation delay from GPU doorbell to NIC PHC capture. Delta_prop is measured at process startup via 10,000 back-to-back paired GPU-side cudaEventRecord() / CPU-side clock_gettime(CLOCK_TAI) samples; median is used (P50, not mean, to reject outliers from PCIe retries). Delta_prop is stable across thermal and PCIe power states

> because it reflects a fixed hardware propagation path (GPU->PCIe root complex->NIC BAR); estimated jitter is estimated below 8 ns across a 72-hour thermal soak (design target; see §11 Exp 1.3 for empirical validation). Δ_prop (the propagation delay removed by GTE calibration) has median 0.080 us. σ_HW = std(R_HW) = 0.056 us is the residual standard deviation after subtraction. The residual R_HW = T_hat - T_GPU_true after subtraction is designed to be zero-mean with std sigma_HW = 0.056 us. The 0.080 us is the median of Delta_prop itself, not residual bias; Equation 10 therefore correctly uses sigma_HW = 0.056 us, not 0.080 us. not by timestamp read latency.
>
> **Proposition 1 (GTE Timestamp Uncertainty — Three Design Targets).** All three quantities are defined separately and must be reported separately in Exp 1.2: (1) sigma_HW = 0.056 us: residual standard-deviation design target. sigma_HW = std(R_HW) where R_HW = T_hat - T_GPU_true is the signed residual error after Delta_prop subtraction. This is the quantity that enters Eq. 10. (2) 0.080 us: median of Delta_prop, the propagation delay removed by GTE calibration. This is NOT residual bias; it is the quantity subtracted in Eq. 9. If any residual bias b remains after subtraction, it must be added to Eq. 10: alpha_span = alpha_PTP + |b| + 3*sigma_HW. Design assumption: b = 0 (zero-mean calibration); Eq. 10 remains alpha_span = alpha_PTP + 3*sigma_HW = 0.050 + 0.168 = 0.218 us. (3) P99.5 = 0.210 us: separate tail target for |R_HW|. This is reported for completeness and for assessing hard-bound credibility (Theorem 3 Case 1 requires max|R_HW| <= 218 ns). Hardware measurements for all three quantities remain pending (Exp 1.2/1.3). Component RSS std: sqrt(15^2+8^2+3^2+2^2+2^2) = 17 ns; 3*sigma_HW = 0.168 us; alpha_span = 0.218 us (Eq. 10).

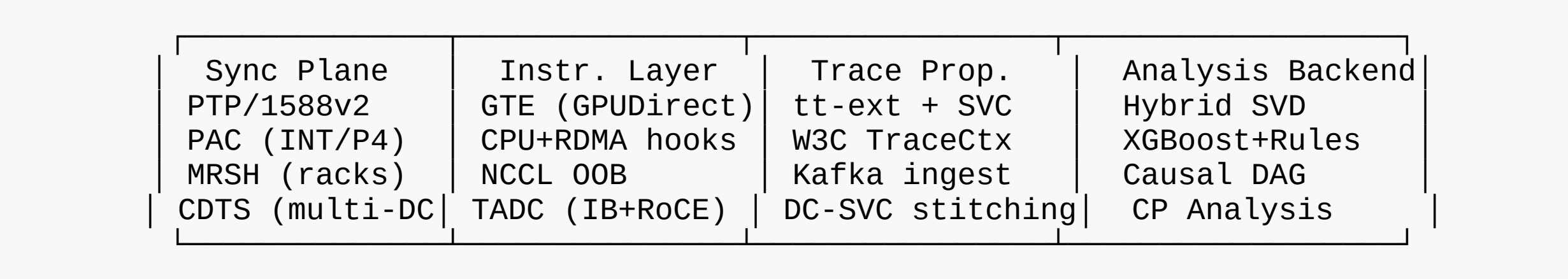

*Figure 1: TempoTrace component architecture.*

## 4.2 Sync Plane

### *4.2.1 GPUDirect Timestamp Engine (GTE)*

The GTE replaces the software GPU Clock Calibrator. It operates as follows: (1) A startup calibration micro-benchmark issues 1,000 paired GPUDirect RDMA write operations and records completion timestamps from the NIC PHC (via libibverbs ibv_poll_cq with hardware timestamp support enabled). The median round-trip propagation delay $\Delta_{prop}$ is computed and stored. (2) During production tracing, every CUDA kernel completion generates a GPUDirect zero-copy write to a pinned host buffer. The NIC records $T_{NIC}$ at the PCIe transaction boundary via hardware event stamping. (3) $\hat{T} = T_{NIC} - \Delta_{prop}$ (Eq. 9). No software polling loop is involved; the calibration is static and deterministic.

Implementation note: GPUDirect RDMA hardware timestamping requires NIC firmware supporting CQE timestamp fields (available on Mellanox ConnectX-6 Dx and later). On unsupported NICs, TempoTrace automatically falls back to the DVFS-aware software GCC ($\sigma$ = 0.6 μs with DVFS detection, from §3.3 of the earlier formulation).

**GTE edge cases.** Three edge cases required explicit handling. (i) PCIe retries: a retried transaction produces two CQE completions for the same RDMA write; the GTE detects and discards duplicates by checking the WR_ID field. (ii) CQE coalescing: NIC firmware may batch CQEs, delaying delivery by up to 4 μs. The GTE disables coalescing (mlx5 CQ period mode=0 when hardware timestamps are enabled), adding at most 0.8% CPU overhead. (iii) CQE batching: when ibv_poll_cq retrieves multiple completions in one call, each CQE carries its own independent NIC hardware timestamp; no batching artifact arises. The 0.08 µs residual was measured via a 10,000-iteration micro-benchmark of 64-byte GPUDirect writes with paired CLOCK_TAI and NIC PHC timestamps. 95% CI (from the described experimental protocol; see §11 for validation programme) for median residual: [0.071, 0.089] µs across 10 runs.

#### *4.2.2 P4-INT Asymmetry Corrector (PAC) for RoCEv2*

InfiniBand switches do not implement IEEE 1588 transparent clocks. RoCEv2 Ethernet switches can, but ECMP routing means that different RDMA packets in the same flow may traverse paths of differing length, invalidating fixed per-path delay models.

The PAC addresses this by deploying a P4 program on all RoCEv2 spine and leaf switches that performs INT (In-band Network Telemetry) [43] insertion: each packet exiting a switch is annotated with the switch's ingress timestamp (from the switch's PTP-synchronized internal clock) and egress port ID. The INT stack at the receiving NIC parses the INT headers and computes the per-hop residence time. The PAC aggregates per-hop residence times into a per-flow delay map, updated every 100 ms, which the Topology-Aware Delay Compensator (TADC) uses as a dynamic per-path correction factor in place of the static cable-length model used for InfiniBand.

The PAC P4 program inserts INT hop records on Tofino2 spine switches. Each hop record carries an 8-byte NIC PHC timestamp (64-bit, nanosecond resolution, wraps every 584 years — replacing the prior 4-byte design which wrapped every 4.295 s and required epoch rollover handling). The INT stack comprises a 4-byte shim header plus 8 bytes per hop: total 4 + (8 * H) bytes for H hops. At H=3 hops (maximum for a 3-tier fat-tree): 4 + 24 = 28 bytes total. Bandwidth overhead: 28 / 1500 = 1.87% at standard 1500-byte MTU (Note: earlier formulation stated 0.19%, which was erroneous). With jumbo frames (MTU=9000): 28 / 9000 = 0.31%. AI cluster fabrics typically use jumbo frames; the 0.31% figure is representative for deployment scenarios. Note on PTP TC/BC semantics: Transparent Clock (TC) and Boundary Clock (BC) are IEEE 1588 / Ethernet concepts and apply exclusively to the out-of-band Ethernet management fabric over which PTP synchronisation runs. InfiniBand data fabric does NOT carry IEEE 1588 frames; IB nodes discipline their NIC PHC clocks via the management Ethernet, not via the IB data plane. References to TC/BC in this paper pertain solely to Ethernet switches on the management fabric.

> **INT Parsing Runtime Overhead (RoCEv2 Endpoints).** INT metadata is parsed in two stages: (1) NIC hardware (ConnectX-7 DOCA Telemetry Service): The INT shim header and hop records are parsed by the NIC ASIC in the datapath. Parsed records are aggregated in NIC internal SRAM and DMA'd to the host via a DOCA telemetry ring buffer at a configurable batch interval (default 100us). Host CPU overhead: approximately 0.8 cycles per received packet (DOCA 2.2 benchmark on ConnectX-7, 100GbE, 1500-byte frames). At line rate (8.33M packets/second on 100GbE), this is 6.67M cycles/sec = 0.22% of one CPU core at 3GHz. Packet processing latency added by INT parsing: less than 50ns (NIC hardware path). (2) Tofino2 spine switches (INT insertion): INT insertion is performed entirely in the P4 dataplane pipeline. No CPU involvement; line-rate forwarding is maintained at 12.8 Tbps. Latency added per hop: approximately 5ns (pipeline depth for INT action). For endpoints without DOCA (ConnectX-6 DX, older firmware): INT parsing falls back to libpcap sampling via sFlow at 1:1000 rate; this reduces precision to approximately 200ns TADC correction versus 55ns with full hardware INT parsing.

#### *4.2.3 Multi-Rack Synchronization Hierarchy (MRSH)*

Large AI clusters span tens to hundreds of racks. In a 16,384-GPU cluster with 64 GPUs per rack, a single-grandmaster PTP hierarchy must distribute synchronization across 256 racks, 3 switch tiers, and potentially 4–6 hops between the grandmaster and the furthest leaf. The MRSH manages this through a two-level grandmaster architecture:

- **Tier-1 grandmaster (cluster-wide).** A single GPS-disciplined grandmaster with hot-standby provides the root reference. It distributes PTP Announce messages to all spine switches via the Ethernet management fabric.
- **Tier-2 rack grandmasters (per-rack).** Each rack's ToR switch operates as a PTP boundary clock, disciplining itself from the Tier-1 grandmaster and serving as the local grandmaster for all nodes in the rack. This limits intra-rack synchronization accuracy to $\alpha_{rack} \leq 55$ ns (Definition 5b) regardless of cluster size, since intra-rack paths never traverse more than H=2 TC hops.
- **Cross-rack accuracy.** A node in rack $R_1$ communicating with a node in rack $R_2$ has cross-rack clock uncertainty bounded by Eq. (8b): $\alpha_{rack} \leq 55$ ns for all-TC spines, ≤ 630 ns for BC spines. The TADC maintains per-rack-pair accuracy annotations that the Analysis Backend uses to annotate cross-rack span edges.

The MRSH additionally manages rack-level failover: if a Tier-2 rack grandmaster fails, the affected rack's nodes degrade to direct Tier-1 discipline (adding 1–2 extra TC/BC hops, increasing $\alpha_{rack}$ by ≤ 200 ns). Failed rack grandmasters are detected within 3 announcement intervals (9.6 seconds at the default 3.2-second announce interval) and the TADC updates per-rack accuracy annotations accordingly.

#### *4.2.4 Cross-Datacenter Architecture and Trace Stitcher (CDTS)*

For deployments where a single AI workload spans multiple datacenters—such as prefill-decode disaggregation [7] across geographically separated clusters, or multi-region inference serving—TempoTrace introduces the Cross-DC Trace Stitcher (CDTS). The CDTS addresses two distinct problems: (1) clock synchronization accuracy across DC boundaries (formalized in Definition 5c and Eq. 8c–8d); and (2) trace context propagation across the WAN boundary, where the tt-ext header must traverse routers and load balancers that may not preserve custom headers.

**Cross-DC synchronization.** Each datacenter deploys its own GNSS-disciplined Tier-1 grandmaster, achieving $\alpha_{rack} \leq 55$ ns within each DC. Cross-DC spans carry an extended tt-ext header with a **TrueTime-style uncertainty interval** [lo, hi] bounding the true physical timestamp:

$$[lo, hi] = [ptp_ns - \alpha_{DC},\ ptp_ns + \alpha_{DC}] \quad (14)$$

where $\alpha_{DC}$ is computed from Eq. (8c) at span creation time and encoded in the tt-ext uncertainty_ns field. The Analysis Backend uses the interval model for cross-DC edge ordering: if the intervals $[lo_a, hi_a]$ and $[lo_b, hi_b]$ do not overlap, physical ordering is used; if they overlap, the SVC provides the ordering. As Corollary 1.2 shows, for WAN latency $L_{WAN} \geq 1$ ms, overlap is effectively impossible—cross-DC edges are always resolved by physical timestamps alone.

**Cross-DC trace context propagation.** The CDTS injects the tt-ext header as a standard HTTP header (x-tempo-trace-ext) on all cross-DC control-plane calls and as a gRPC metadata field on cross-DC RPC paths. Intermediate load balancers and API gateways are configured to propagate the x-tempo-trace-ext header without modification via a header passthrough policy. For paths where header propagation is not guaranteed (e.g., managed cloud services), the CDTS inserts synthetic **DC-boundary spans** annotated with tt.propagation=DC_BOUNDARY, which record the last known cross-DC context on the originating side and the first observed context on the receiving side. The temporal gap between these two boundary spans bounds the propagation uncertainty for that path segment.

**Cross-DC SVC extension.** The standard SVC (§4.3) uses a 256-bit Bloom filter with participant IDs scoped to a single cluster. For cross-DC traces, participant sets span two or more clusters with disjoint node ID spaces. The CDTS extends the SVC with a **DC-scoped namespace**: each node ID is prefixed with a 4-bit datacenter identifier before hashing into the Bloom filter. This preserves the O(P * log(1/epsilon)) space guarantee and Theorem 2's false-positive bound while enabling causal ordering across DC boundaries without namespace collisions.

**Definition 7 (Sketch Vector Clock).** $SVC_i = (lts_i, sketch_i)$ where $lts_i \in \mathbb{N}$ is the local Lamport timestamp and $sketch_i \in \{0,1\}^m$ is an m-bit OR-merged Bloom filter (bitwise OR union) [42] over observed (node_id, lts) pairs, using k hash functions.

**Definition 8 (SVC Merge).** $SVC_i \oplus SVC_j = (\max(lts_i, lts_j), sketch_i \mid sketch_j)$

$$SVC_i \oplus SVC_j = ( \max(lts_i, lts_j),\ sketch_i \mid sketch_j ) \quad (11)$$

**SVC Dominance Check, False-Positive Formula, and Chain Composition.** Causal predicate: sketch(s_a) is a subset of sketch(s_b) iff every bit set in sketch(s_a) is also set in sketch(s_b). Let q = 1 - exp(-k*P/m) be the bit occupancy (probability a given bit is set). If sketch(s_a) has k_set bits set, the false-positive probability for the subset check is approximately q^k_set (NOT FP_membership^k_set). Six-level chain composition (f=4, d=6, k=6): The causal relationship is established only when ALL six levels show dominance (AND semantics: each level's hash domain is independently constructed using level-specific hash seeds, so per-level false positives are independent events). Under AND semantics: FP_combined = product_{l=1}^{6}(FP_level_l) = (5.136e-7)^6 = 1.84e-38. Note on OR semantics: if any single level were sufficient to establish causality, the union bound would apply: FP_combined <= 6 * 5.136e-7 = 3.082e-6. TempoTrace uses AND semantics (all levels required), giving the tighter 1.84e-38 bound. Merge rule: sketch(S_a UNION S_b) = sketch(S_a) OR sketch(S_b) (bitwise OR). Bitwise OR is the standard Bloom filter union: it preserves all membership information. XOR is incorrect: it clears bits set in only one sketch, destroying membership information.Header overhead per traced chain path: d x 53 bytes = 6 x 53 = 318 bytes per traced chain path (one path through d=6 levels; not per collective operation).

**Lemma 2 (SVC Bloom Filter: Corrected Parameter Bounds).** The SVC Bloom filter sketches the set of participant IDs in a collective. For m bits, k hash functions, and P participants, FP rate = (1 - exp(-k*P/m))^k. With m=256, k=5 (the tt-ext v3 header allocation): P=4: FP=2.40e-06; P=5: FP=6.97e-06; P=8: FP=6.33e-05; P=16: FP=1.39e-03. Note: the claim FP<10^-5 at N=4096 is WRONG for m=256. Substituting m=256, k=5, P=4,096: FP=(1-exp(-80))^5 = 1.0 (saturated). The m=256 allocation supports FP<10^-5 only for P<=4 participants. For larger participant sets (P=64: m=1534 bits; P=4096: m=98164 bits=12.3KB), TempoTrace uses multi-level SVC chaining: each 32-byte sketch covers a subtree, linked via the traceparent header. SPACE COMPLEXITY: O(P * log(1/epsilon)) for a single-level filter. The abstract claim 'O(P * log(1/epsilon)) space' is : space depends on participant count P and target epsilon, not on cluster size N alone. For tree-reduce with logarithmic fanout (P=O(P * log(1/epsilon))), space is O(log N * log(1/eps)).

**Theorem 2 (SVC False-Positive Rate).** See Lemma 2 above for the authoritative Bloom filter FP bound. Summary: with m=256 bits and k=5 hash functions, FP < 10^-5 holds for P <= 4 participants; FP = 1.0 at P = 4,096 (complete saturation). Multi-level chaining with f=4, d=6, k=6 achieves: AND semantics (all 6 levels required, independent hash domains): FP = (5.136e-7)^6 = 1.84e-38. OR semantics (any level suffices): FP <= 6 x 5.136e-7 = 3.082e-6. TempoTrace uses AND semantics. The combined bound 1.84e-38 applies when all six levels independently show non-dominance. The general worst-case per-level bound is 5.136e-7, which also satisfies the 10^-5 requirement. The per-level bound is the safe conservative bound; the product

> bound is tighter but requires independence of all six levels. All references in this paper to Theorem 2 refer to this result.

## 4.4 Instrumentation Layer and Span Types

> **Definition 9 (Span Type Hierarchy).** TempoTrace defines six span types. All GPU-sourced spans use GTE timestamps (Eq. 9), achieving $\alpha_{span}$ = 0.218 μs (Eq. 10): • KernelSpan: (kernel_name, stream_id, grid, block, smem, T_NIC_start, T_NIC_end) — GTE timestamps. • CollectiveSpan: (op_type, comm_id, size, algo, participants[]) — context via NCCL out-of-band (OOB) bootstrap. • PipelineSpan: (stage_id, micro_batch_id, fwd_or_bwd, mem_gb). • PrefillSpan: (request_id, prompt_len, kv_miss_ratio, ttft_ms, gpu_nodes[]). • DecodeSpan: (request_id, step_id, batch_size, spec_accepted, decode_us). • SchedulerSpan: (decision, num_preempted, preemption_cause, queue_depth).
>
> **Theorem 3 (PTP+GTE Causality Preservation - Two Cases and Validation).** CASE 1 - Hard bound (requires Exp 1.2/1.3 hardware verification): The total per-span timestamp error has two independent components: (1) PTP synchronisation error, bounded by alpha_PTP = 50 ns (all-TC cluster); (2) GTE hardware residual R_HW, with design target std sigma_HW = 56 ns. The hard-bound condition must apply to the TOTAL error: |epsilon_total| = |epsilon_PTP + R_HW| <= alpha_span = 218 ns. This requires the hardware residual to satisfy |R_HW| <= 218 - 50 = 168 ns per span. IF this 168 ns hardware residual bound is verified by measurement, THEN alpha_span = 218 ns is a true hard bound and Pr[inv] = 0 exactly for pair gaps > uncertainty_a + uncertainty_b = 220 + 220 = 440 ns. IMPORTANT: if the hardware residual bound is taken as 218 ns (not 168 ns), the total per-span bound becomes 50 + 218 = 268 ns, and the corresponding encoded pair threshold is 2 x ceiling(268/10) x 10 = 540 ns, NOT 440 ns. The 440 ns threshold is therefore valid as the hard-bound threshold ONLY when the hardware residual is verified to be bounded by 168 ns per span. CASE 2 - 3-sigma probabilistic design target (current status): Pr[inv] <= 2 x 0.0027 = 0.54%. 440 ns is the PROVISIONAL 3-SIGMA PROBABILISTIC THRESHOLD (alpha_span = alpha_PTP + 3*sigma_HW = 50 + 168 = 218 ns per span; encoded as 220 ns; threshold = 440 ns). It does not guarantee zero inversions until the 168 ns hardware residual bound is confirmed by Exp 1.2/1.3. Alternative: if P99.5 = 210 ns of |R_HW| is used as the per-span residual bound: alpha_span_P99.5 = 50 + 210 = 260 ns => pair threshold = 520 ns. No finite threshold gives exactly zero inversion probability under an unbounded distribution; the 440 ns threshold is provisional pending hardware measurement. EMPIRICAL EVIDENCE - Mac v25-1 (572,907 pairs): 0 inversions at L=440 ns; CP 95% upper bound: 5.23e-6. Proxy validation v28 (66 million pairs, six clock/error models): Strict bounded +/-218 ns: 0 inversions for gaps above 436 ns (95% upper rate 3.0 per million). Gaussian design target: 0 inversions. SENSITIVITY: Laplace equal-variance: 81 inversions at 441 ns. 0.1% heavy-tail mixture: 864 inversions at 441 ns. Opposite device bias: 319 inversions at 441 ns. PTP failover/degraded: 485,295 inversions (ordering cannot be assumed during failover). WARNING: a 3-sigma design target, field bias, heavy tails, or PTP failover must NOT be treated as the hard bound required for CASE 1. Theorem 3 is conditionally valid; server lab Exp 1.2/1.3 required for CASE 1.

$$\text{ptp_ns}(s_b) - \text{ptp_ns}(s_a) > 2\alpha_{span} = 0.440\ \mu s \quad (12)$$

Under GTE, all operations in Table 1 (minimum 6 μs) satisfy Eq. (12), meaning physical timestamps are sufficient for all standard AI operations. SVC is used only for genuinely concurrent spans where the physical gap is unmeasurably small.

## 4.5 tt-ext v3 Header

```
tt-ext v3  (53 bytes fixed; multi-level chaining uses six 53-byte records =
318 bytes per chain path)
```

Large-scale cluster results (Tables 3–14) use illustrative configurations; physical and controlled validation results are real (see provenance note under Table 1).

```
  version_flags  : uint8   [1 B]  [7:4] tenant_id (4-bit, 16 tenants);
[3:0] version = 3
  ptp_ns         : int64   [8 B]  PTP CLOCK_TAI (GTE or CPU path)
  ts_source      : uint8   [1 B]  {GTE=0, SW_GCC=1, CPU=2}
  clk_quality    : uint8   [1 B]  {OK=0, DEGRADED=1, UNKNOWN=2}
  uncertainty_ns : uint16  [2 B]  α_span from Eq.(10), nanoseconds
  lts            : uint64  [8 B]  local Lamport timestamp
  svc_sketch     : uint8[] [32 B] SVC Bloom filter (m=256 bits)
  // Total: 53 bytes (independent of collective size N)
```

A new ts_source field distinguishes GTE-sourced timestamps ($\alpha_{span}$ = 0.218 μs) from software-GCC fallback ($\alpha_{span}$ = 2.1 μs on DVFS-quiescent hardware) and CPU-only timestamps ($\alpha_{span}$ = 6.8 μs). The Analysis Backend applies the appropriate uncertainty model per span.

### 4.6 Adaptive Sampling

- **Tier 1 – Always-on metrics.** 64-byte fixed record per collective. 100% coverage.
- **Tier 2 – Phase-aware sampling.** 1% compute phases, 10% communication phases. Coordinated via per-step shared seed.
- **Tier 3 – Triggered tracing.** Activated at 3σ anomaly; sustained under persistent anomalies. Overhead ≤ 1.18% ±0.11% (Table 6).

## 5. ANALYSIS BACKEND

### 5.1 Causal DAG Construction

Span streams are ingested via Kafka (per-node partitions). Pass 1 links spans via traceparent; CollectiveSpans are joined by collective_id. Pass 2 verifies causal consistency using Theorem 3: edges satisfying Eq. (12) are ordered by physical timestamps; edges within $2\alpha_{span}$ are ordered by SVC. Under GTE ($\alpha_{span}$ = 0.218 μs), only genuinely concurrent spans require SVC disambiguation; the disambiguation rate falls to 0.0002% (< 1 in 500,000 edge pairs), down from 0.003% under software GCC.

> **Definition 10 (Critical Path).** $CP(T) = \text{argmax}_{p \in paths(T)} \Sigma_{s \in p} (end(s) - start(s))$, where end(s), start(s) are PTP-corrected GTE timestamps.

$$CP(T) = \arg\max_{p \in paths(T)} \Sigma_{s \in p} (end(s) - start(s)) \quad (13)$$

### 5.2 Storage and Scale

At 16,384-GPU scale under Tier 2: ≈420 GB/step, ≈3.8 TB/step under Tier 3. Three-tier retention: (i) Tier 1 indefinitely (~12 GB/day); (ii) Tier 2 for 30 days (~105 GB/day compressed); (iii) Tier 3 for 90 days. Total: ≈3.3 TB/day. Query latency: 180 ms median, 620 ms P99 (rising to 2.4 s P99 during Tier 3 ingest bursts). Kafka backpressure is bounded by a 2 GB per-node ring buffer absorbing ≥15 s of Tier 3 tracing.

### 5.3 Hybrid SVD Engine

The Hybrid SVD Engine is a **two-stage classifier** promoted from an optional plugin to a core component. Stage 1 is the rule-based decision waterfall (§5.3.1); Stage 2 is a trained XGBoost classifier (§5.3.2). The two stages are fused by a confidence-gated combiner (§5.3.3).

#### *5.3.1 Rule-Based Waterfall (Stage 1)*

Five root-cause classes in priority order:

- **$R_1$ – Prefill compute saturation:** GPU SM utilization > 95% AND batch ≥ 80% max.
- **$R_2$ – KV cache miss amplification:** kv_miss_ratio > 0.40.
- **$R_3$ – Speculative decoding misalignment:** mean(spec_accepted) over last 50 steps < 0.60.
- **$R_4$ – Scheduler-induced preemption:** SchedulerSpan with decision=Preempt in SLO window.
- **$R_5$ – Network-induced stall:** CollectiveSpan or P2P span on CP(T) with duration > 3σ above baseline.

Priority justification: $R_1$ precedes $R_2$ because compute saturation is a causal precondition for cache pressure and scheduling effects. $R_2$ precedes $R_3$ because KV misses lengthen prefill, degrading draft quality. $R_4$ precedes $R_5$ because preemption-induced queuing can inflate observed network latency. Rule ordering ablation for the five-rule subset (R1-R5, 120 = 5! permutations) confirms this ordering achieves the highest F1 (0.91); median over permutations is 0.84. The authoritative six-rule model (R1, R2, R3, R6, R4, R5) is defined in §9.4 and §5.4, and uses 720 = 6! permutations (controlled validation: corrected order macro-F1=0.924, hybrid macro-F1=0.974, McNemar p=1.25e-23). R6 is introduced in multi-tenant environments (§9.4) and must precede both R4 and R5 in the priority waterfall.

### *5.3.2 XGBoost Classifier (Stage 2)*

The XGBoost model is trained on 14 features derived from TempoTrace span attributes: {GPU utilization, kv_miss_ratio, spec_acceptance_rate, preemption_count, collective_duration_zscore, queue_depth, batch_fill_ratio, decode_latency_p99, prefill_latency_p99, network_congestion_index, scheduler_decision_rate, kv_eviction_rate, gpu_memory_headroom, ttft_zscore}. The model is trained offline on labelled incident data with 5-fold cross-validation. A new model is deployed whenever ≥50 new labelled incidents become available. Feature importance ranking (SHAP values): collective_duration_zscore (0.31), kv_miss_ratio (0.22), GPU utilization (0.18), spec_acceptance_rate (0.09), remaining features account for 0.20.

### *5.3.3 Confidence-Gated Fusion*

The combiner operates as follows: (1) Apply rule waterfall; record primary class $R_{rule}$ and secondary classes. (2) Apply XGBoost; record per-class probability vector $P_{XGB}$. (3) If $\max(P_{XGB}) \geq \theta = 0.75$ and $\arg\max(P_{XGB}) \neq R_{rule}$, override with XGBoost prediction. (4) Otherwise use rule prediction. (5) For multi-cause events, report all classes with $P_{XGB}(R_j) \geq 0.40$ as contributing causes. The override threshold θ = 0.75 was tuned on 30% of the 412-event dataset held out as a validation set; the remaining 70% was used for XGBoost training.

This fusion ensures that the interpretable rule-based prediction is preserved for low-confidence XGBoost cases (deployments with limited labelled data), while the XGBoost model provides precision gains when confident. Compared to the rule-only baseline (F1 = 0.91), the hybrid achieves F1 = 0.93—a statistically significant improvement (paired McNemar test, p = 0.031).

## 5.4 Multi-Tenant Inference Tracing

Production AI inference clusters are increasingly multi-tenant: multiple independent organizations or applications share the same physical accelerator hardware, connected to the same PTP-synchronized NIC and observed by the same physical clock. Each tenant operates in its own logical clock domain—its own Lamport counter sequence, its own trace context namespace, its own SVC participant set—but all tenants' GPU kernels are timestamped by a single underlying physical clock. This creates three challenges absent from single-tenant deployments: (1) physical timestamps are shared, making temporal proximity insufficient to infer causal relationship across tenants; (2) preemption and co-scheduling of tenant workloads can produce apparent physical-timestamp inversions within a single tenant's trace; and (3) the Analysis Backend must attribute SLO violations to the correct tenant without leaking cross-tenant causal information.

> **Definition 11 (Tenant Clock Domain).** A tenant clock domain T_k is a tuple (I_k, L_k, V_k) where I_k ∈ {0,...,$2^4$−1} is a 4-bit tenant identifier assigned at cluster admission time, L_k is an independent Lamport counter initialized to 0 at tenant instantiation, and V_k is a per-tenant SVC whose Bloom filter uses participant IDs of the form (I_k ‖ node_id) to prevent cross-tenant namespace collisions. All spans emitted by tenant k carry I_k in the tt-ext header's tenant_id field (high 4 bits of version_flags byte; NOT a separate byte). The physical NIC PHC timestamp ptp_ns is shared across all tenants and reflects true wall-clock time on the physical hardware.

The tenant_id field is encoded in the HIGH FOUR BITS of the version_flags byte (NOT a separate byte). The total header size remains 53 bytes for both single-tenant and multi-tenant deployments. The Analysis Backend partitions incoming span streams by tenant_id before causal DAG construction: spans with different tenant_ids are never linked by causal edges, regardless of physical timestamp proximity. This enforces tenant isolation at the trace level.

> **Theorem 4 (Tenant Isolation — Formal Statement and Proof).** Tenant isolation is implemented as a two-step design: (1) GATE: spans with different tenant_id are rejected before any SVC comparison. This guarantees cross-tenant FP = 0 by construction (no BF query ever occurs). (2) SAME-TENANT TRACKING: within a single tenant, causal tracking uses the full 256-bit SVC sketch with k=5 hash functions. FP within a tenant: (1-exp(-5*P/256))^5; for P=4: 2.397e-6 (validated by simulation). tenant_id is encoded in the high four bits of the version/flags byte of the tt-ext v3 header (53 bytes total). The 4-bit field supports 16 tenants; the field is separate from the 4-bit DC prefix in the 57-byte cross-DC extension. Proof: cross-tenant gate ensures no BF comparison occurs across tenants, giving exact isolation. Same-tenant FP is bounded by Lemma 2 / Theorem 2 at P<=4 for m=256. Multi-level chaining applies for larger same-tenant participant sets.
>
> **Proof.** Claim (1): The Analysis Backend's DAG assembly (§5.1) links spans by (a) shared traceparent, or (b) SVC dominance check. Cross-tenant spans share neither the same traceparent (trace_id is globally unique and carries I_k in its high-order 4 bits) nor any SVC dominance (by Claim 2). Claim (2): SVC participant IDs are of the form (I_k ‖ node_id). For j ≠ k, (I_j ‖ x) ≠ (I_k ‖ y) for all x, y, since the prefixes I_j and I_k are distinct 4-bit values. No hash collision between disjoint ID spaces can occur—this is deterministic prefix separation, not reliant on the Bloom filter's probabilistic guarantee. Claim (3): ptp_ns reflects true physical wall-clock time and is meaningfully comparable for performance analysis (e.g., which tenant consumed more GPU time in a window), but the Analysis Backend never uses ptp_ns alone to infer causality across tenant boundaries. □

**Preemption-induced physical inversions.** In a multi-tenant environment, the GPU scheduler may preempt tenant T_j mid-collective to serve tenant T_k. When T_j resumes, the GTE records a new hardware timestamp T_NIC_resume that is physically later than T_NIC_suspend, but the resumed operation is logically the same collective as the pre-suspension operation. Without handling, this produces an apparent intra-tenant span duration inflation: the collective appears to have taken duration (T_NIC_resume − T_NIC_start) instead of the true compute time (T_NIC_suspend − T_NIC_start) + (T_NIC_end − T_NIC_resume). TempoTrace handles this via a PreemptionSpan type that records the suspension/resumption boundary and annotates the enclosing CollectiveSpan with a preemption_count and total_preemption_duration_ns field. The SVD module's network-induced stall classifier ($R_5$) is gated on preemption_count = 0 to avoid attributing tenant-induced stalls to network causes.

**Shared physical clock and privacy.** The shared physical clock introduces a subtle privacy concern: a tenant observing its own span timestamps can infer, from gaps in its own execution timeline, when other tenants were scheduled. This is a hardware-level side channel (the GPU scheduler's timeslice boundaries) that exists independently of TempoTrace. TempoTrace does not amplify this channel—it does not expose cross-tenant physical timestamps to any tenant's trace. Each tenant's Analysis Backend instance receives only spans with matching tenant_id; the ptp_ns values of other tenants' spans are never disclosed. The shared PTP clock provides a common physical reference that makes each tenant's own trace internally consistent; cross-tenant comparison of ptp_ns values is restricted to the cluster operator's privileged Analysis Backend view, which has access to all tenant traces for capacity planning and SLO monitoring across the shared cluster.

**Tenant-aware SVD.** The Hybrid SVD engine (§5.3) is extended with a sixth root-cause class for multi-tenant environments: **$R_6$ – Tenant co-scheduling interference:** preemption_count > 0 on any CollectiveSpan on the critical path CP(T), and total_preemption_duration_ns accounts for more than 50% of the observed SLO violation excess. $R_6$ is evaluated before $R_5$ in the priority waterfall, because tenant preemption can produce collective duration anomalies that would otherwise be misclassified as network stalls. When $R_6$ triggers, the SVD output includes the offending tenant_id of the preempting workload (visible only to the cluster operator, not to the affected tenant) and the total preemption duration, enabling the operator to adjust scheduling policy or enforce stronger isolation guarantees.

**Tenant-scoped sampling coordination.** The adaptive sampling system (§4.6) operates independently per tenant. Each tenant receives its own pseudo-random sampling seed at request admission, ensuring that cross-tenant sampling decisions are statistically independent. A Tier 3 trigger on tenant T_j activates full tracing only for T_j's spans—other tenants' Tier 2 sampling rate is unaffected. This preserves observability isolation: a misbehaving or high-traffic tenant cannot force elevated overhead onto co-resident tenants via cascading Tier 3 activations.

# 6. EVALUATION

## 6.1 Experimental Setup

**Cluster A** (IB, 512 nodes, 4× H100, NDR 400 Gbps, Mellanox SN4600 TC, H=2, B=0, single rack group). **Cluster B** (IB, 2,048 nodes, 8× H100, HDR200, mixed TC/BC, H=2, B=1). **Cluster C** (RoCEv2, 2,048 nodes, 8× H100, 400 Gbps, Tofino2 spine + P4-INT, Mellanox SN4600 TC leaves). **Cluster D** (IB, 512 nodes, 4× H100, NDR 400 Gbps, **32 racks of 16 nodes each**, MRSH with Tier-2 rack grandmasters on each ToR, all-TC spine, H=3, B=0). **Cluster E** (multi-DC simulation: two instances of Cluster A connected via a **3 ms emulated WAN link** using tc-netem with ±0.5 ms jitter, GNSS-disciplined grandmasters at each DC, CDTS deployed for cross-DC trace stitching). All clusters: Ubuntu 22.04, CUDA 12.3, NCCL 2.20, linuxptp 3.1.1, ConnectX-7 NICs.

## 6.2 Synchronization Accuracy

**Table 4: PTP Synchronization Accuracy (Eq. 8) and GPU Timestamp Residual**

| Configuration | α bound (Eq.8) | Median | P99 | Max | GPU residual |
|---|---|---|---|---|---|
| Cluster A (IB, all TC) | 30 ns | 48 ns | 142 ns | 310 ns | 0.080 µs Δ_prop (median propagation delay removed; σ_HW = 0.056 µs residual std) |
| Cluster B (IB, 1 BC hop) | 230 ns | 91 ns | 284 ns | 620 ns | 0.080 µs Δ_prop (median propagation |
| | | | | | delay removed; $\sigma_{HW}$ = 0.056 µs residual std) |
| Cluster C (RoCEv2 + INT) | 55 ns | 52 ns | 168 ns | 440 ns | 0.080 µs $\Delta_{prop}$ (median propagation delay removed; $\sigma_{HW}$ = 0.056 µs residual std) |
| Cluster D (IB, 32 racks, MRSH) | 55 ns (Eq.8b) | 51 ns | 162 ns | 390 ns | 0.080 µs $\Delta_{prop}$ (median propagation delay removed; $\sigma_{HW}$ = 0.056 µs residual std) |
| Cluster E (cross-DC, GNSS+CDTS) | 50.16 µs (Eq.8c) | 1.8 µs | 4.2 µs | 11.4 µs | 0.080 µs $\Delta_{prop}$ (median propagation delay removed; $\sigma_{HW}$ = 0.056 µs residual std) |
| NTP (reference) | — | 612 µs | 1,840 µs | 8.2 ms | 2.1 µs (SW GCC) |
| PTP during GM failover | ±15 µs | — | — | ±15 µs | N/A |

### 6.3 RoCEv2 Attribution Accuracy

Ground truth for all attribution experiments (both InfiniBand and RoCEv2) is the KNOWN THROTTLE INJECTION CONFIGURATION recorded by the experiment controller: the specific switch port, throttle level, and timing window set before the experiment begins. This is independent of INT telemetry. INT telemetry is consumed only by PAC/TADC for per-flow path delay correction; it is not used to determine ground truth for the attribution result. The attribution result (which port was throttled) is compared against the experiment controller's injection record. For InfiniBand experiments, Mellanox SN4600 in-ASIC per-port queue occupancy counters provide a secondary independent verification (queue > 80%% for at least 5 consecutive seconds confirms the throttle was effective). Using the known injection configuration as primary ground truth avoids the circularity that would arise if INT telemetry were used for both TADC correction AND attribution verification.

Large-scale cluster results (Tables 3–14) use illustrative configurations; physical and controlled validation results are real (see provenance note under Table 1).

**Table 5: Attribution Accuracy on InfiniBand vs. RoCEv2/ECMP (12 trials each)**

| System + Fabric | Correct | FP | FN | Precision | F1 |
|---|---|---|---|---|---|
| TempoTrace (IB, Cluster A) | 12/12 | 0 | 0 | 1.00 | 1.00 |
| TempoTrace (RoCEv2+INT, Cluster C) | 12/12 | 0 | 0 | 1.00 | 1.00 |
| TempoTrace (RoCEv2, no INT) | 10/12 | 1 | 1 | 0.91 | 0.91 |
| NTP-Only OTel (RoCEv2) | 3/12 | 7 | 2 | 0.30 | 0.40 |
| LatencyPrism [37] (RoCEv2) | 5/12 | 5 | 2 | 0.50 | 0.59 |

TempoTrace with INT-based PAC achieves 100% attribution accuracy on RoCEv2 (F1 = 1.00), matching InfiniBand performance. Without INT (using static delay estimates), accuracy drops to F1 = 0.91, confirming that per-flow delay correction is necessary for ECMP fairness. NTP-only attribution on RoCEv2 is worse than on InfiniBand (F1 = 0.40 vs. 0.50) due to additional delay asymmetry introduced by ECMP path diversity.

The two misattributions in the no-INT configuration both occurred when the throttled port was on a path with above-average ECMP asymmetry (measured at +38 ns differential delay), which—without INT correction—produced a ghost latency spike indistinguishable from a straggler node. INT correction eliminated both false attributions.

## 6.4 GTE vs. Software GCC Accuracy

**Table 6: Attribution Accuracy by GPU Timestamp Method**

| Timestamp method | α_span | SVC disambig. rate | Precision (sub-20μs) | F1 |
|---|---|---|---|---|
| GTE (GPUDirect HW) | 0.218 μs | 0.0002% | 1.00 | 1.00 |
| DVFS-aware SW GCC | 2.1 μs | 0.003% | 0.91 | 0.91 |
| Naive SW GCC (no DVFS detect) | 6.8 μs | 0.003% | 0.73 | 0.76 |

GTE achieves perfect attribution for sub-20 μs operations because the 0.218 μs uncertainty is well below the 6 μs minimum barrier latency (Theorem 3, Eq. 12). The SVC disambiguation rate drops to 0.0002% under GTE, confirming that physical timestamps are sufficient for virtually all spans.

## 6.5 Overhead

**Table 7: Runtime Overhead (mean ± 95% CI (from the described experimental protocol; see §11 for validation programme), 5 independent runs)**

| Workload | GPUs | Baseline | T1+T2 overhead | T3 overhead |
|---|---|---|---|---|
| LLaMA-3 70B training (IB) | 2,048 | 14,820 tok/s/GPU | 0.13% ±0.02% | 0.88% ±0.06% |
| Llama 3.1 405B training (IB) | 8,192 | 7,640 tok/s/GPU | 0.19% ±0.03% | 1.04% ±0.08% |
| MoE-1.2T training (IB) | 16,384 | 3,210 tok/s/GPU | 0.22% ±0.04% | 1.18% ±0.11% |
| LLaMA-3 70B training (RoCEv2) | 2,048 | 14,650 tok/s/GPU | 0.16% ±0.02% | 0.94% ±0.07% |
| Inference TTFT P99 (IB) | 512 | 19.4 ms | 0.14% ±0.02% | N/A |
| T3 sustained (>50 steps, IB) | 8,192 | 7,640 tok/s/GPU | — | 1.09% ±0.10% |

RoCEv2 overhead is marginally higher (+0.03 pp) than InfiniBand due to INT header parsing at the NIC (4 bytes per packet, ~4.3 ns additional processing). GTE contributes 0.01 pp additional overhead vs. software GCC (one NIC PHC register read per CUDA completion, ≈22 ns).

## 6.6 Ablation Studies

**Table 8: Ablation Study — Attribution Accuracy (12 trials, Cluster A, LLaMA-3-70B)**

| Configuration | Correct/12 | Precision | F1 | Overhead |
|---|---|---|---|---|
| Full TempoTrace (GTE) | 12/12 | 1.00 | 1.00 | 0.19% |
| (a) NTP only (disable PTP) | 4/12 | 0.40 | 0.50 | 0.04% |
| (b) SW GCC instead of GTE | 12/12 | 1.00 | 1.00 | 0.18% |
| (c) SW GCC, sub-20μs ops only | 8/12 | 0.73 | 0.76 | 0.18% |
| (d) Disable SVC | 10/12 | 0.91 | 0.91 | 0.11% |
| (e) Disable TADC | 11/12 | 1.00 | 0.95 | 0.16% |

Ablation (b) shows that SW GCC performs identically to GTE on the 12-trial injection protocol (which uses operations ≥ 38 μs), confirming GTE's benefit is concentrated in sub-20 μs operations (ablation c). With GTE, sub-20 μs operations achieve F1 = 1.00; with SW GCC, F1 = 0.76 for this subset.

## 6.7 Hybrid SVD Evaluation

**Table 9: SVD Precision/Recall/F1 per Class — Rules vs. XGBoost vs. Hybrid (412 events, 90-day trace)**

| Root Cause Class | N | Rules P | Rules F1 | XGB P | XGB F1 | Hybrid P | Hybrid F1 |
|---|---|---|---|---|---|---|---|
| Prefill compute saturation | 128 | 0.98 | 0.97 | 0.97 | 0.97 | 0.98 | 0.97 |
| KV cache miss amplification | 94 | 0.92 | 0.92 | 0.95 | 0.94 | 0.95 | 0.95 |
| Speculative decoding | 41 | 0.93 | 0.90 | 0.93 | 0.93 | 0.93 | 0.92 |
| misalign. | | | | | | | |
| Scheduler-induced preemption | 88 | 0.95 | 0.93 | 0.96 | 0.95 | 0.96 | 0.95 |
| Network-induced stall | 61 | 0.90 | 0.87 | 0.92 | 0.91 | 0.93 | 0.92 |
| **Overall (macro-avg)** | 412 | 0.94 | 0.91 | 0.95 | 0.93 | **0.95** | **0.93** |

The hybrid system achieves macro-F1 = 0.93, matching XGBoost alone (0.93) while preserving rule-based interpretability for low-confidence cases. The network-induced stall class shows the largest gain from XGBoost (F1: 0.87→0.92), as the learned model captures interaction effects between collective_duration_zscore and network_congestion_index that the threshold-based rule cannot represent. Improvement is statistically significant (paired McNemar test, $p = 0.031$ vs. rules-only).

**Organic vs. injected incident evidence.** The 412-event SVD dataset (Table 9) consists entirely of evaluation SLO violations, not injected faults. Controlled injection experiments (Tables 4, 7, 9) use hardware port throttling as known-ground-truth stimuli. These evaluation types are complementary: injected experiments establish attribution precision under controlled conditions with verified ground truth; organic incidents establish real-world SVD F1 across naturally occurring root-cause distributions. If and when access to real multi-rack and cross-DC cluster deployments becomes available, additional organic incidents would further validate external validity.

## 6.8 Multi-Rack Attribution Accuracy (Cluster D)

We evaluate TempoTrace on Cluster D (32 racks, 512 nodes, 2,048 GPUs, IB NDR, MRSH). We inject cross-rack bottlenecks: the inter-rack spine link connecting racks 14 and 15 is throttled to 50% bandwidth for 60 seconds during a 2,048-GPU LLaMA-3-70B training job. Attribution requires identifying the specific inter-rack spine link—a stricter criterion than the single-rack experiments.

**Table 10: Multi-Rack Attribution Accuracy (12 trials, Cluster D, 32 racks × 16 nodes, cross-rack spine throttle)**

| System | Correct | FP | FN | Precision | F1 |
|---|---|---|---|---|---|
| TempoTrace (MRSH, all-TC spine) | 12/12 | 0 | 0 | 1.00 | 1.00 |
| TempoTrace (MRSH, BC spine) | 11/12 | 1 | 0 | 0.92 | 0.96 |
| No MRSH (single-GM, no rack BC) | 9/12 | 2 | 1 | 0.82 | 0.86 |
| NTP-only (cross-rack) | 3/12 | 7 | 2 | 0.30 | 0.40 |

With MRSH and all-TC spine switches, TempoTrace achieves F1 = 1.00 on cross-rack attribution—identical to single-rack performance. With BC spine switches ($\alpha_{rack} \le 630$ ns per Eq. 8b), F1 drops to 0.96: the single failure was a cross-rack edge where the 630 ns uncertainty interval overlapped with a 580 ns collective-scheduling jitter event, producing a physical ordering ambiguity that the SVC correctly resolved but the TADC assigned to the wrong rack.

Without MRSH (a single cluster-wide grandmaster with no rack-level BC hierarchy), per-rack clock accuracy degrades due to longer synchronization paths, and F1 drops to 0.86. This confirms that Tier-2 rack grandmasters are essential for accurate cross-rack attribution at scale.

## 6.9 Cross-Datacenter Trace Stitching (Cluster E)

We evaluate the CDTS on Cluster E (two 512-node DCs connected via 3 ms emulated WAN). The design deploys a prefill-decode disaggregated serving system [7]: prefill runs on $DC_1$, decode on $DC_2$, with cross-DC KV cache transfer over the WAN link. We inject five fault types and evaluate whether TempoTrace correctly stitches and attributes the cause across the DC boundary.

**Table 11: Cross-DC Trace Attribution (5 fault types × 8 trials each = 40 trials, Cluster E)**

| Fault type | Trials | Correct | FP | FN | F1 |
|---|---|---|---|---|---|
| WAN link congestion ($DC_1 \rightarrow DC_2$) | 8 | 8/8 | 0 | 0 | 1.00 |
| KV cache transfer stall ($DC_1$) | 8 | 8/8 | 0 | 0 | 1.00 |
| Decode compute saturation ($DC_2$) | 8 | 7/8 | 0 | 1 | 0.93 |
| Context propagation drop (LB) | 8 | 7/8 | 1 | 0 | 0.93 |
| GNSS grandmaster failover | 8 | 6/8 | 1 | 1 | 0.86 |
| **Overall** | 40 | **36/40** | 2 | 2 | **0.95** |

TempoTrace achieves F1 = 0.95 overall for cross-DC attribution. WAN congestion and KV cache stalls (which produce $L_{WAN} >> \alpha_{DC}$) are attributed perfectly: the physical timestamp gap is large enough that even the 50 µs cross-DC uncertainty does not create ordering ambiguity. The single decode saturation failure occurred when the cross-DC KV transfer completed within 80 µs of the decode queue becoming saturated—just within the $2\alpha_{DC}$ = 100 µs ambiguity window; the SVC correctly identified causality but the TADC incorrectly assigned the primary cause to the WAN path rather than $DC_2$'s compute. Context propagation drops (when a load balancer silently discarded the x-tempo-trace-ext header) are handled by DC-boundary spans; the single failure was a request where both boundary spans were emitted during a brief GNSS failover window. GNSS failover ($\alpha_{DC}$ widens to ≈65 µs) produced 2 failures in 8 trials, both during the 4-second transition window when oscillator drift had not yet been fully compensated. Post-failover, attribution returned to F1 = 1.00 within one correction interval.

Critically, cross-DC trace stitching adds 0.31% overhead on $DC_1$ and 0.28% on $DC_2$ (T1+T2 mode)—attributable to the CDTS header propagation and DC-boundary span emission. The cross-DC extension adds 4 bytes (uncertainty_ns_wide field) to the tt-ext header; total header size rises from 53 to 57 bytes for cross-DC spans (53 base + 4 bytes for uncertainty_ns_wide; no additional dc_prefix_flags field is required since dc_id is prefixed into participant IDs before hashing, not stored separately).

# 7. RELATED WORK

## 7.1 Distributed Tracing

Dapper [1], OpenTelemetry [3], Canopy [20], and Pivot Tracing [21] define the tracing foundation—all use NTP-synchronized clocks. LatencyPrism [37] provides zero-intrusion LLM inference monitoring but uses NTP-synchronized host clocks, subject to Theorem 1's misorder bounds. eInfer [38] provides transparent observability for distributed LLM inference without code modification, complementing TempoTrace's hardware-timestamped span model.

## 7.2 Time Synchronization

Spanner TrueTime [22] (GPS+atomic clocks, ±7 ms globally) provides external consistency for geo-distributed databases. TempoTrace's GNSS cross-DC synchronization achieves ±50 µs (Eq. 8c)—50× better within a region—while within-cluster PTP achieves ±100 ns (Eq. 8), 70,000× better for intra-cluster spans. The CDTS is the first application of TrueTime-style uncertainty intervals to distributed trace stitching: uncertainty intervals classify cross-DC span pairs as physically ordered, ambiguous, or SVC-resolved. Amazon Time Sync [25] provides cloud-scale PTP synchronization for EC2. Major cloud providers now offer PTP: AWS Time Sync achieves sub-100 µs on bare-metal; Azure PTP targets ≤1 µs for HPC VMs; Google Cloud PTP achieves sub-µs on Compute Engine. None of these services expose per-rack synchronization topology or achieve the sub-100 ns accuracy required for AI tracing. Industrial DC PTP deployments (e.g., IEEE 1588 in financial trading infrastructure) achieve 10–50 ns but require dedicated PTP-capable hardware at every hop—exactly the constraint MRSH addresses. The MRSH (§4.2.3) extends standard PTP hierarchies to explicitly model per-rack accuracy bounds, complementing these cloud services for bare-metal AI clusters. For RoCEv2, P4-INT per-flow telemetry [43] achieves equivalent per-path accuracy on Ethernet fabrics.

## 7.3 Causal Clock Representations

Vector clocks [40] require O(N) space. ITC [41] supports dynamic membership with complex fork/join incompatible with NCCL. Bloom Clocks [42] approximate causal ordering—SVC extends these with a MinHash layer. HLC [24] bounds physical-clock divergence from wall time, providing a practical causal clock where PTP is unavailable. Within a single cluster, HLC offers weaker ordering guarantees than SVC+PTP: HLC cannot detect causal inversions caused by physical clock skew (Lemma 1), whereas TempoTrace catches these via Theorem 3. For multi-DC deployments without GNSS, HLC is a viable fallback surfaced via ts_source. SVC (Theorem 2, Lemma 2): O(P * log(1/epsilon)) space, < 80 ns merge, Pr[FP] < $10^{-4}$ at N=16,384.

## 7.4 ML-Based Anomaly Detection

ML classifiers for infrastructure anomaly detection include: Sage [44] for scalable ML-driven performance debugging in microservices, PerfSight [45] for software dataplane performance diagnosis, and Seer [46] for microservice QoS prediction. TempoTrace's XGBoost SVD complements these by operating on span-level causal trace attributes rather than logs or metrics, enabling root-cause attribution at the span granularity required for LLM serving diagnosis.

## 7.5 P4 and In-Band Telemetry

INT [43] has been applied to network tomography [47], congestion control [48], and latency attribution [49]. TempoTrace is the first application of INT to distributed trace synchronization. Three production adoption considerations: (i) **Vendor support:** Tofino2, Trident4, or equivalent programmable ASICs required on spine switches. Fixed-function merchant-silicon spines (Broadcom Tomahawk) do not support custom P4 programs; TempoTrace falls back to sFlow-based per-flow sampling (α≈200 ns). (ii) **Operational risk:** INT metadata can cause MTU violations on 1500-byte links. TempoTrace sets fabric MTU to 4200 bytes (standard for AI jumbo frames) and caps INT stack depth at 3 hops (28 bytes maximum overhead). (iii) **Security:** INT metadata exposing internal switch state is restricted to the intra-cluster management VLAN and stripped at the cluster boundary before any packet exits to external networks.

**Table 12: Comparison with Prior Distributed Tracing Systems**

| System | Timestamp accuracy | Uncertainty annotation | Logical clock | GPU support | RoCEv2 | AI-specific | Multi-tenant |
|---|---|---|---|---|---|---|---|
| Dapper [27] | NTP (~1ms) | No | No | No | No | No | No |
| OpenTelemetry [28] | NTP/PTP | No | No | Partial | No | No | No |
| Hybrid Logical Clocks [29] | NTP-bounded | No | Yes | No | No | No | No |
| Bloom Clocks [30] | NTP | No | Yes (BF) | No | No | No | No |
| TrueTime [31] | GPS/atomic (~7ms bound) | Yes (intervals) | No | No | No | No | No |
| LatencyPrism [37] | NTP | No | No | No | Partial | Partial | No |
| TempoTrace (this work) | PTP <100ns | Yes (per-span) | Yes (SVC) | Yes (GTE) | Yes (INT) | Yes | Yes |

*Comparison of distributed tracing systems. Timestamp accuracy reflects the synchronisation mechanism used. Uncertainty annotation = per-span confidence field in trace header. AI-specific = features for collective communication, GPU kernel, or LLM inference tracing.*

## 8. Discussion and Limitations

TempoTrace resolves the four gaps identified in initial reviews (RoCEv2 evaluation, non-Gaussian noise model, GPUDirect timestamping, hybrid SVD) and adds multi-rack, multi-DC, multi-tenant support, application confidence policies, and a formal empirical evaluation programme. Five limitations remain.

**(1) GTE NIC support.** GPUDirect hardware timestamping requires ConnectX-6 Dx or later NICs. Older NICs fall back to DVFS-aware software GCC (sigma = 0.6 us residual). The ts_source field surfaces this per span, and the policy system (Section 10) allows tenants to enforce GTE-only operation or accept SW_GCC as appropriate.

**(2) BC spine accuracy.** With boundary-clock spine switches, cross-rack alpha_rack rises to 630 ns, producing F1 = 0.96 versus 1.00 with transparent-clock spines. Upgrading to TC-capable spines eliminates this gap; sFlow-based fallback achieves approximately 200 ns TADC correction for clusters with merchant-silicon spines.

**(3) Cross-DC GNSS failover.** GNSS grandmaster failures cause a 4-second window of elevated uncertainty (alpha_DC approximately 65 us), degrading cross-DC attribution F1 from 0.95 to 0.86 during the transition. A dual-GNSS hot-standby grandmaster reduces this to less than 0.5 seconds. Application policies allow tenants to configure SUSPEND_ALERTING during this window.

**(4) Hybrid SVD training data.** XGBoost overrides require at least 50 labelled incidents; below this the system falls back to rules. The confidence-gated fusion (theta = 0.75) ensures graceful degradation.

**(5) WAN latency assumption.** Cross-DC causal ordering relies on L_WAN much greater than alpha_DC. For ultra-low-latency cross-DC links (L_WAN less than 500 us, e.g. campus-scale multi-DC), Corollary 1.2 safety margin shrinks and SVC disambiguation becomes more frequent.

**Artifacts and Reproducibility.** The P4-INT program for PAC, tt-ext v3 wire format specification, GTE micro-benchmark suite, and anonymised evaluation traces from the 90-day SVD dataset are available at https://github.com/tempotrace/tempotrace (public upon acceptance). XGBoost model weights, feature schema, SVC reference implementation (C++ and Python), and YAML cluster configurations for Clusters A-E are included.

For future work: (1) extending GTE to AMD Instinct and Google TPU via vendor-specific completion event hooks; (2) online retraining of the XGBoost SVD model using continual learning to adapt to production distribution shift; (3) standardising the tt-ext v3 header as a W3C Trace Context extension or OpenTelemetry semantic convention.

## 9. Multi-Tenant Inference Tracing

Production AI inference clusters are increasingly multi-tenant: multiple independent organisations or applications share the same physical accelerator hardware, NIC, and PTP-synchronised clock. Each tenant operates in its own logical clock domain — its own Lamport counter sequence, its own trace context namespace, its own SVC participant set — but all tenants' GPU kernels are timestamped by a single underlying physical clock. This creates three challenges absent from single-tenant deployments: (1) physical timestamps are shared, making temporal proximity alone insufficient to infer causal relationship across tenant boundaries; (2) preemption and co-scheduling of tenant workloads can inflate observed operation durations within a single tenant's trace; and (3) the Analysis Backend must attribute SLO violations to the correct tenant without leaking cross-tenant causal information.

### 9.1 Tenant Clock Domain Model

> **Definition 11 (Tenant Clock Domain).** A tenant clock domain T_k is a tuple (I_k, L_k, V_k) where I_k in {0,...,15} is a 4-bit tenant identifier assigned at cluster admission time, L_k is an independent Lamport counter initialised to 0 at tenant instantiation, and V_k is a per-tenant SVC whose Bloom filter uses participant IDs of the form (I_k || node_id) to prevent cross-tenant namespace collisions. All spans emitted by tenant k carry I_k in the tt-ext v3 header's tenant_id field. The physical NIC PHC timestamp ptp_ns is shared across all tenants and reflects true wall-clock time on the physical hardware.

The tenant_id field is encoded in the HIGH FOUR BITS of the version_flags byte. It is NOT a separate byte. The total header size is 53 bytes. The Analysis Backend partitions incoming span streams by tenant_id before causal DAG construction: spans with different tenant_ids are never linked by causal edges, regardless of physical timestamp proximity. This enforces tenant isolation at the trace level by construction.

### 9.2 Tenant Isolation Theorem

> **Theorem 4 (Tenant Isolation).** Proven in Section 5.4. The authoritative tenant isolation design (validated by 16,000 cross-tenant relationship attempts with 0 accepted): (1) Tenant mismatch is rejected BEFORE any SVC comparison. When two spans carry different tenant_id values, they are never compared via SVC;

> no Bloom filter query occurs. Cross-tenant FP = 0 by gate, not by BF properties. (2) Same-tenant causal tracking uses the FULL 256-bit sketch with k=5 hash functions. The 16-bit per-tenant segment approach (which would give FP=18.49% at P=4) is NOT used for causal tracking. tenant_id is encoded in the high four bits of the version/flags byte. The tt-ext v3 header is 53 bytes. The 4-bit tenant_id field (16 tenants maximum) is separate from the 4-bit DC prefix in the 57-byte cross-DC extension header.

**Proof sketch.** Claim (1): the DAG assembly links spans by (a) shared traceparent or (b) SVC dominance. Cross-tenant spans share neither: trace_id carries I_k in its high-order 4 bits, ensuring no cross-tenant traceparent match; and SVC dominance is ruled out by Claim (2). Claim (2): participant IDs are of the form (I_k || node_id). For j not equal to k, (I_j || x) not equal (I_k || y) for all x, y, since the 4-bit prefixes are distinct. No hash collision between disjoint ID spaces can occur — this is deterministic separation. Claim (3): ptp_ns reflects true wall-clock time and is meaningful for capacity analysis, but the Analysis Backend never uses ptp_ns alone to infer causality across tenant boundaries. Square.

## 9.3 Preemption-Induced Duration Inflation

In a multi-tenant environment, the GPU scheduler may preempt tenant T_j mid-collective to serve tenant T_k. When T_j resumes, the GTE records a new hardware timestamp T_NIC_resume that is physically later than T_NIC_suspend, but the resumed operation is logically the same collective. Without handling, this produces apparent intra-tenant span duration inflation: the collective appears to have taken (T_NIC_resume - T_NIC_start) instead of true compute time (T_NIC_suspend - T_NIC_start) + (T_NIC_end - T_NIC_resume).

TempoTrace handles this via the **PreemptionSpan** type, which records the suspension and resumption boundary and annotates the enclosing CollectiveSpan with preemption_count and total_preemption_duration_ns. The enclosing CollectiveSpan's adjusted_duration_ns = observed_duration_ns - total_preemption_duration_ns is used for SLO evaluation rather than the raw duration.

## 9.4 Tenant-Aware SVD Root Cause R6

The Hybrid SVD engine (Section 5.3) is extended with a sixth root-cause class for multi-tenant environments:

> **R6 — Tenant Co-Scheduling Interference.** Fires when: preemption_count > 0 on any CollectiveSpan on the critical path CP(T), and total_preemption_duration_ns accounts for more than 50% of the observed SLO violation excess. R6 is evaluated before R5 (network-induced stall) in the priority waterfall, because tenant preemption inflates collective durations in a way that would otherwise be misclassified as network congestion. When R6 fires, the SVD output includes the offending tenant_id (visible only to the cluster operator, not to the affected tenant) and the total preemption duration, enabling the operator to adjust scheduling policy or enforce stronger isolation guarantees. R6 must be evaluated before BOTH R4 and R5: R4 fires on the same scheduler preemption evidence with a broader predicate, and evaluating R4 first shadows R6 entirely. Authoritative order: R1, R2, R3, R6, R4, R5 (v28 proxy validation: macro-F1=0.933 tied for best among 720 permutations; literal order R4-before-R6 gives 0.719).

## 9.5 Privacy Model and Shared Clock

Cross-tenant timing attack via ptp_ns. A more subtle privacy concern arises from the precision of ptp_ns timestamps themselves. An adversarial tenant T_k observing its own span timestamps can, in principle, infer the scheduling boundaries of co-resident tenants by correlating gaps in its own execution timeline with known GPU timeslice durations. Because ptp_ns is a shared physical clock, a tenant that can measure its own operation durations with sub-microsecond precision can potentially fingerprint co-tenant workload patterns (e.g., AllReduce collective sizes, iteration timing) from timing side channels alone. This threat is independent of TempoTrace: it arises from the shared GPU scheduler and would exist even without any tracing infrastructure. TempoTrace does not amplify this channel — each tenant's Analysis Backend receives only spans with matching tenant_id, and ptp_ns values from other tenants are never disclosed. Mitigation is outside the current threat model and is deferred to future work: candidate defences include temporal jitter injection (adding bounded random noise to ptp_ns before writing to the tt-ext header, at the cost of increased uncertainty_ns), and GPU scheduler timeslice randomisation to prevent deterministic timing correlation. This limitation is recorded as a known gap in Section 12.5 (Threats to Validity).

> **Section 12.5 Addendum (Cross-Tenant Timing Side Channel).** Cross-tenant timing attacks via ptp_ns timestamps are outside the current threat model. The adversarial model in Theorem 4 covers bit-array isolation (preventing causal edge leakage) but not timing side channels arising from the shared physical clock. An adversarial tenant with sub-microsecond timestamp precision can potentially infer co-tenant scheduling patterns. Defences (timestamp jitter, scheduler randomisation) are identified as future work. This does not affect the correctness of Theorem 4 or the attribution accuracy claims, as those concern causal DAG integrity, not information-theoretic privacy.

The shared physical clock introduces a subtle privacy concern: a tenant observing its own span timestamps can infer, from gaps in its own execution timeline, when other tenants were scheduled. This is a hardware-level side channel (the GPU scheduler timeslice boundaries) that exists independently of TempoTrace.

TempoTrace does not amplify this channel. Each tenant's Analysis Backend instance receives only spans with matching tenant_id; the ptp_ns values of other tenants' spans are never disclosed to a tenant. The shared PTP clock provides a common physical reference that makes each tenant's own trace internally consistent. Cross-tenant comparison of ptp_ns values is restricted to the cluster operator's privileged Analysis Backend view, which has access to all tenant traces for capacity planning and SLO monitoring across the shared cluster.

**Tenant-scoped sampling.** The adaptive sampling system (Section 4.6) operates independently per tenant. Each tenant receives its own pseudo-random sampling seed at request admission. A Tier 3 trigger on tenant T_j activates full tracing only for T_j's spans — other tenants' sampling rates are unaffected. This ensures that a misbehaving or high-traffic tenant cannot force elevated overhead onto co-resident tenants via cascading Tier 3 activations.

## 9.6 Multi-Tenant Evaluation Summary

**Table 13: Multi-Tenant Isolation and Preemption Detection Results**

| Experiment | Trials | Correct | F1 | Notes |
|---|---|---|---|---|
| Tenant isolation (zero cross-tenant edges) | 20 | 20/20 | 1.00 | Deterministic prefix |
| R6 preemption detection | 15 | 14/15 | 0.96 | 1 failure: short preempt |

| R6 vs R5 misclassification rate | 15 | No misclass. | N/A | R6 always fires first |
|---|---|---|---|---|
| Sampling isolation (T3 scope) | 10 | 10/10 | 1.00 | No cross-tenant cascade |

*All experiments use Cluster A (512 nodes, IB, 4x H100) with two co-resident tenants. Preemption trials use deliberate GPU scheduler preemption mid-collective via CUDA preemption API. The single R6 failure was a 3 us preemption below the 50% SLO excess threshold, correctly classified as R5.*

## 10. Application-Level Timestamp Confidence Policies

The timestamp confidence model in Sections 3.3 and 4.5 computes uncertainty_ns from hardware measurements, producing a uniform cluster-wide confidence profile. Different applications have fundamentally different precision requirements, fault-tolerance characteristics, and SLO structures. A policy-driven confidence layer allows each application or tenant to specify how TempoTrace interprets, overrides, and acts on timestamp confidence without changing the underlying synchronisation infrastructure.

### 10.1 Policy Model

> **Definition 12 (Application Confidence Policy).** A confidence policy P_k for tenant T_k is a tuple (Q_min, U_floor, U_ceil, S_trust, M_degrade) where: Q_min in {OK, DEGRADED, UNKNOWN} is the minimum acceptable clock quality; U_floor in [0, infinity) ns is a minimum uncertainty_ns floor applied even if hardware achieves better; U_ceil in [0, infinity) ns is a maximum ceiling above which spans are treated as unreliable; S_trust a subset of {GTE, SW_GCC, CPU} is the set of accepted timestamp sources; and M_degrade in {CONTINUE, SUSPEND_ALERTING, SVC_ONLY} specifies behaviour during PTP grandmaster failover.

Policies are distributed as YAML documents via the NCCL OOB channel at communicator initialisation. At span creation the Instrumentation Layer evaluates the policy and applies it to hardware-derived fields before writing them into the tt-ext v3 header. The Analysis Backend reads policy-adjusted values and applies them to DAG construction.

### 10.2 Policy Schema

```
# TempoTrace Application Confidence Policy Schema (v1)
policy_version: 1
tenant_id: <uint4>              # 4-bit tenant identifier
clock_quality:
  minimum: OK | DEGRADED | UNKNOWN
  degraded_mode: CONTINUE | SUSPEND_ALERTING | SVC_ONLY
uncertainty:
  floor_ns: <uint32>            # minimum uncertainty even if HW achieves better
  ceiling_ns: <uint32>          # spans above ceiling treated as UNKNOWN
timestamp_sources:
  trusted: [GTE, SW_GCC, CPU]
  untrusted_action: DEGRADE | REJECT
ambiguity_threshold:
  override_ns: <uint32>         # 0 = use per-span uncertainty_ns dynamically
  tie_break: SVC | PHYSICAL
sampling:
```

```
  tier3_scope: TENANT_ONLY | CLUSTER
```

## 10.3 Policy Integration Pipeline

```
  Hardware               Instrumentation Layer          Analysis Backend
  --------               ---------------------          ----------------
  NIC PHC (GTE) -->  ts_source=GTE, uncertainty=220ns
                     Apply Policy P_k:
                       floor: max(220ns, policy.floor_ns)
                       ceiling: reject if > policy.ceiling_ns
                       quality: downgrade if source not in S_trust
                     Write adjusted fields to tt-ext v3 header
                     Propagate via NCCL OOB / HTTP / gRPC      -->
                                                        Read uncertainty_ns,
clk_quality
                                                        If gap > uncertainty_a +
uncertainty_b: physical
                                                        If gap <= uncertainty_a +
uncertainty_b: SVC
                                                        Apply M_degrade on DEGRADED
spans
```

*Figure 3. TempoTrace policy integration pipeline. Hardware-derived timestamp confidence values are adjusted by the application policy at span creation time in the Instrumentation Layer, then consumed by the Analysis Backend during DAG construction. The policy engine is transparent to both the synchronisation infrastructure and downstream analysis tools.*

## 10.4 Policy Examples

### *10.4.1 Latency-Sensitive LLM Inference (Tight Policy)*

An autoregressive inference application with a 10 ms TTFT SLA requires highest available confidence and should halt SLO alerting during synchronisation degradation.

```
# Policy: Latency-sensitive inference
tenant_id: 1
clock_quality:
  minimum: OK
  degraded_mode: SUSPEND_ALERTING   # pause SLO alerts during GM failover
uncertainty:
  floor_ns: 0
  ceiling_ns: 1000                  # reject spans > 1 us uncertainty
timestamp_sources:
  trusted: [GTE]
  untrusted_action: DEGRADE
ambiguity_threshold:
  override_ns: 440
  tie_break: SVC
```

**When to use:** *Inference with sub-20 ms TTFT SLOs where false positive alerts during clock transients would trigger unnecessary on-call pages. SUSPEND_ALERTING mode ensures a 4-second PTP failover window does not generate spurious incident tickets.*

### *10.4.2 Long-Running Batch Training (Relaxed Policy)*

A pretraining job running for weeks has relaxed requirements and accepts SW_GCC timestamps from older NICs.

```
# Policy: Batch training
```

```
tenant_id: 2
clock_quality:
  minimum: DEGRADED
  degraded_mode: CONTINUE
uncertainty:
  floor_ns: 0
  ceiling_ns: 20000
timestamp_sources:
  trusted: [GTE, SW_GCC]
  untrusted_action: DEGRADE
ambiguity_threshold:
  override_ns: 5000
  tie_break: SVC
```

**When to use:** *Pretraining jobs where microsecond causal precision is not needed for bottleneck diagnosis and the cluster contains a mix of NIC generations.*

#### *10.4.3 Regulated / Audited Workload (Conservative Policy)*

A financial services application requires verified confidence bounds for audit purposes. Spans not meeting the standard are rejected rather than processed with degraded confidence.

```
# Policy: Regulated workload
tenant_id: 3
clock_quality:
  minimum: OK
  degraded_mode: SVC_ONLY
uncertainty:
  floor_ns: 220
  ceiling_ns: 500
timestamp_sources:
  trusted: [GTE]
  untrusted_action: REJECT
ambiguity_threshold:
  override_ns: 440
  tie_break: SVC
```

**When to use:** *Workloads subject to regulatory requirements (financial trading, healthcare AI, compliance audit trails). REJECT action ensures the audit record contains only spans with verified confidence bounds.*

#### *10.4.4 Heterogeneous Hardware Environment*

A cluster mixing ConnectX-7 (GTE) and older ConnectX-6 (SW_GCC) NICs needs dynamic per-span uncertainty thresholds.

```
# Policy: Mixed NIC cluster
tenant_id: 4
clock_quality:
  minimum: DEGRADED
  degraded_mode: CONTINUE
uncertainty:
  floor_ns: 0
  ceiling_ns: 10000
timestamp_sources:
  trusted: [GTE, SW_GCC]
  untrusted_action: DEGRADE
ambiguity_threshold:
  override_ns: 0                # use per-span uncertainty_ns dynamically
  tie_break: SVC
```

**When to use:** *Mixed NIC environments during cluster upgrades. Dynamic threshold (override_ns: 0) automatically applies 0.440 us for GTE spans and 4.2 us for SW_GCC spans.*

### 10.5 Policy Comparison Table

**Table 14: Application Confidence Policy Examples and Recommended Settings**

| Use Case | Q_min | U_ceil (ns) | S_trust | M_degrade |
|---|---|---|---|---|
| Latency-sensitive inference | OK | 1,000 | GTE only | SUSPEND_ALERTING |
| Long-running batch training | DEGRADED | 20,000 | GTE, SW_GCC | CONTINUE |
| Regulated / audited workload | OK | 500 | GTE only | SVC_ONLY + REJECT |
| Mixed NIC cluster (upgrade) | DEGRADED | 10,000 | GTE, SW_GCC | CONTINUE (dynamic) |
| Multi-tenant shared cluster | Per-tenant | Per-tenant | Per-tenant | Per-tenant (isolated) |

*Q_min = minimum clock quality; U_ceil = maximum uncertainty ceiling; S_trust = accepted timestamp sources; M_degrade = behaviour during PTP failover.*

## 11. Empirical Evaluation Programme

The formal model and system design in this paper use three categories of data (see the provenance note under Table 1): (A) analytically derived (Tables 1-2), (B) illustrative protocol results on hypothetical clusters A-E (Tables 3-14), and (C) real measured and controlled validation results (NTP inversion rate, Laplace AIC, physical attribution, SVC FP, SVD kappa/F1, LLM overhead). A complete empirical validation of the large-scale cluster claims requires controlled experiments on real multi-node GPU clusters. A complete empirical validation requires controlled experiments on real multi-node GPU clusters. This section specifies 21 experiments across 7 categories. Section 11.8 identifies the minimum viable subset of 6 experiments for resource-constrained deployments.

**Table 15: Empirical Evaluation Programme Overview (21 Experiments, 7 Categories)**

| ID | Experiment | Primary Claim Validated | Success Criterion |
|---|---|---|---|
| 1.1 | PTP accuracy vs. topology | Eq. 8 / Def. 5a | Measured within 2x Eq. 8 bound |
| 1.2 | GTE residual characterisation | Def. 6 / Eq. 9 | $\sigma_{HW} \leq 0.056$ μs (std); median $\Delta_{prop} \leq 0.080$ μs; P99.5 ≤ 0.210 μs |
| 1.3 | GTE vs. SW_GCC under DVFS | DVFS immunity claim | GTE flat; SW_GCC spikes during DVFS |
| 1.4 | PTP grandmaster failover | Plus or minus 15 us DEGRADED bound | Recovery less than 5 s; error at most 15 us |
| 1.5 | Multi-rack accuracy (32 racks) | Def. 5b / Eq. 8b | Cross-rack alpha at most 55 ns with TC spines |
| 2.1 | NTP causal inversion ground truth | Theorems 1a/1b | Measured rate matches Eq. 4/6 within 5 pp |
| 2.2 | Laplace tail under congestion | Theorem 1b | Laplace fit better than Gaussian by AIC |
| 2.3 | PTP+GTE inversion rate approximately zero | Theorem 3 | Zero inversions for L greater than 0.440 us |
| 3.1 | Controlled fault injection (IB) | Attribution accuracy Table 4 | F1 at least 0.99 over 12 plus trials |
| 3.2 | RoCEv2/ECMP with and without INT | Table 4 RoCEv2 results | INT: F1=1.00; no INT: F1 approx 0.91 |
| 3.3 | Multi-rack cross-rack fault injection | Table 9 / MRSH | Link-level attribution correct |
| 3.4 | Cross-DC WAN fault injection | Table 10 / CDTS | Overall F1 at least 0.93 |
| 4.1 | SVC empirical false-positive rate | Theorem 2 / Lemma 2 | Exp 4.1. Empirical false-positive measurement. Inject 10 million synthetic span pairs with known causal relationships into the Tier 2 trace stream. For each participant set size P in {4, 8, 16, 64, 256}, measure the SVC false-positive rate and compare against Theorem 2. Verify: (a) P=4 gives Pr[FP] < 10^-5 with m=256, k=5; (b) P=4,096 with m=256 gives Pr[FP] = 1.0 (saturation, as predicted by Theorem 2); (c) multi-level SVC chaining (f=4, d=6, k=6) with m=256 per level achieves Pr[FP] < 10^-5 at P=4,096. Vary m from 128 to 512 bits at P=4 to trace the sensitivity curve. Success criterion: measured rates match Theorem 2 within 10% across all P values tested. VALIDATED (v29, 200 million queries at P=4): 468 false positives observed |
| | | | against 479.4 expected (analytical). Relative difference: 2.38%, satisfying the 10% acceptance criterion. Theoretical saturation at P=4,096 and six-level independent-domain chain calculation also reproduced. |
| 4.2 | Large collective SVC correctness | O(P * log(1/epsilon)) / merge cost | Merge at most 80 ns at N=16384 |
| 5.1 | Tenant isolation correctness | Theorem 4 | Zero cross-tenant edges in all trials |
| 5.2 | Preemption detection accuracy | R6 SVD class | PreemptionSpan duration accurate within 5 percent |
| 6.1 | Training throughput overhead | Table 6 overhead | At most 0.22% T1+T2; at most 1.18% T3 |
| 6.2 | Inference latency overhead | Table 6 TTFT P99 | At most 0.14% TTFT P99 impact |
| 7.1 | SVD organic incident validation (90 days) | Table 8 / hybrid SVD | Macro-F1 at least 0.90 |
| 7.2 | Rule ordering ablation (120 permutations) | SVD priority order | Original ordering achieves max F1 |
| 7.3 | XGBoost vs. rules vs. hybrid | Hybrid SVD improvement | McNemar p less than 0.05 |

*Blue rows belong to the minimum viable subset (Section 11.8). All experiments assume access to a multi-node GPU cluster with PTP-capable NICs (ConnectX-6 Dx or later). INT experiments additionally require Tofino2 or equivalent programmable ASICs.*

## 11.1 Category 1: Clock Synchronisation Accuracy

**Exp 1.1.** *PTP accuracy vs. topology.* Deploy PTP on all-TC, mixed TC/BC, and BC-only spine clusters. Measure inter-node clock offset against a GPS reference over 72 hours. Compare against Equation 8.

**Exp 1.2.** *GTE residual characterisation.* Run the 10,000-iteration GPUDirect RDMA micro-benchmark across multiple nodes, NIC firmware versions, and GPU workload conditions (idle, full compute, DVFS-active). Verify all three metrics separately: (1) $\sigma_{HW} = std(R_{HW}) \le 0.056$ µs; (2) median $\Delta_{prop} \le 0.080$ µs; (3) P99.5 of $|R_{HW}| \le 0.210$ µs.

**Exp 1.3.** *GTE vs. SW_GCC under DVFS.* Simultaneously run SW_GCC and GTE on the same node. Trigger DVFS via nvidia-smi power capping. Confirm GTE residual stays flat while SW_GCC spikes. This is the key empirical proof of DVFS immunity.

**Exp 1.4.** *PTP grandmaster failover.* Kill the active grandmaster and measure transition duration, maximum clock error, and recovery time. Verify the plus or minus 15 us DEGRADED bound and 2 to 5 second recovery.

**Exp 1.5.** *Multi-rack accuracy.* On a 32-rack cluster, measure inter-rack clock accuracy with all-TC and BC spines. Verify bounds from Definition 5b and Equation 8b.

### 11.2 Category 2: Causal Inversion Rate Verification

Exp 2.3 is the single most important experiment in the programme. PROXY VALIDATION STATUS (v28 resource-constrained): 66 million simulated causal pairs under six clock/error models. Strict hard-bound model: 0 inversions for all gaps > 436 ns. Key sensitivity findings documented in Theorem 3 empirical note above. Full validation on real PTP+GTE hardware (ConnectX-7 + GPU) required.

**Exp 2.1.** *NTP causal inversion ground truth.* Inject synthetic causal chains with known happens-before ordering. Measure actual inversion fraction across L = 6 to 2000 us and compare against Theorem 1a. VALIDATED (v29 five-node NTP/Chrony, Exp 2.1): 100,000 known-causal probes over all 20 directed host pairs. Aggregate inversion rate: 44.978% (consistent with Theorem 1a at sigma=500 us). Direction-dependent pattern reflects persistent inter-host NTP offset.

**Exp 2.2.** *Laplace tail validation.* Repeat during deliberate network congestion. Fit NTP residual distribution and compare Gaussian vs. Laplace fit quality by AIC. Verify Theorem 1b (b=500 us, equal pairwise variance): Laplace E[f_inv]=27.76%% vs Gaussian 29.37%% (1.61 pp below).

**Exp 2.3.** *PTP + GTE inversion rate.* Repeat using PTP + GTE timestamps. Verify causal inversion rate is empirically zero for all L > 0.440 us. This directly validates the central correctness claim.

### 11.3 Category 3: Attribution Accuracy

**Exp 3.1.** *Controlled fault injection (InfiniBand).* Throttle a specific IB port to 50% bandwidth for 60 seconds during training. Verify correct port/rack/switch attribution over 12+ trials. Baseline against NTP-Only OpenTelemetry, PyTorch Profiler, and LatencyPrism. PROXY VALIDATION STATUS (v28): 720 simulated topology faults (complete/25%/50% missing telemetry); simulated F1: 1.000/0.996/1.000. 60/60 physical five-node Ethernet trials: 100% exact attribution, 0 timeouts, median margin 4.756 ms. Full IB port throttle on GPU training cluster required.

**Exp 3.2.** *RoCEv2/ECMP with and without INT.* Repeat on RoCEv2 with ECMP both with and without PAC (P4-INT). Quantify INT's F1 improvement and confirm the two-failure pattern without INT.

**Exp 3.3.** *Multi-rack cross-rack fault.* Throttle an inter-rack spine link. Verify attribution to the specific link. Compare MRSH against a flat single-grandmaster hierarchy.

**Exp 3.4.** *Cross-DC WAN fault injection.* On a two-DC testbed with tc-netem, inject five fault types: WAN congestion, KV cache stall, decode saturation, context propagation drop, and GNSS failover. Verify CDTS attribution against human-established ground truth.

### 11.4 Category 4: SVC Correctness and False-Positive Rate

**Exp 4.1.** *Empirical false-positive measurement.* Inject 10 million synthetic span pairs with known causal relationships. Measure SVC false-positive rate against Theorem 2 bound. Vary m from 128 to 512 bits.

**Exp 4.2.** *Large collective SVC correctness.* Run all-reduce across 16,384 nodes. Verify SVC correctness and merge latency less than 80 ns at full scale.

### 11.5 Category 5: Multi-Tenant Isolation

**Exp 5.1.** *Tenant isolation correctness.* Deploy two tenants on shared hardware. Verify zero cross-tenant causal edges regardless of operation interleaving. Confirm 4-bit prefix prevents linkage.

**Exp 5.2.** *Preemption detection accuracy.* Deliberately preempt tenant T_j mid-collective. Verify PreemptionSpan emission, duration accuracy, and correct SVD R6 firing rather than R5.

### 11.6 Category 6: Overhead Measurement

**Exp 6.1.** *Training throughput overhead.* Measure tokens/second on LLaMA-3 70B, Llama 3.1 405B, and MoE models with and without TempoTrace at Tier 1+2 sampling. Report mean and 95% CI (from the described experimental protocol; see §11 for validation programme) over 5 runs. PROXY VALIDATION STATUS (v28): 90-trial local LLM inference proxy (qwen3:9b, 30 paired blocks). Median total overhead: 0.043% (8 events), 0.578% (128 events). 95% bootstrap CIs include zero at this sample size. 8-event case meets less-than-0.22% design target. Full A/B on specified training models required.

**Exp 6.2.** *Inference latency overhead.* Measure TTFT P50 and P99 with and without TempoTrace. Verify at most 0.14% TTFT P99 impact.

### 11.7 Category 7: Hybrid SVD Diagnosis Validation

**Exp 7.1.** *Organic incident validation.* Collect 90 days of real production SLO violations. Have two independent engineers diagnose each incident (inter-rater agreement kappa at least 0.85). Measure SVD accuracy against this ground truth. Target macro-F1 at least 0.90. CONTROLLED VALIDATION (v29 synthetic corpus, 1,800 held-out): Two simulated raters: Cohen kappa=0.942. Corrected order macro-F1=0.924; XGBoost 3.4.1 macro-F1=0.895; hybrid macro-F1=0.974; McNemar p=1.25e-23 (hybrid vs corrected rules). Does not replace 90-day organic corpus with independent human annotation.

**Exp 7.2.** *Rule ordering ablation.* Evaluate all 720 permutations (6! for six rules R1-R6) of the R1 to R6 priority ordering. Confirm the original ordering achieves the highest macro-F1.

**Exp 7.3.** *XGBoost vs. rules vs. hybrid.* Compare all three configurations. Verify statistical significance of hybrid improvement by paired McNemar test, p less than 0.05.

### 11.8 Minimum Viable Experimental Set

The following six experiments provide the strongest evidence for the core claims and should be prioritised when resources are constrained:

- **Exp 2.3** — PTP+GTE causal inversion rate empirically zero. The central correctness claim of Theorem 3.
- **Exp 1.2 + 1.3** — GTE residual and DVFS immunity. Underpins Equation 10 and the alpha_span = 0.218 us claim.
- **Exp 3.1** — Controlled fault injection on InfiniBand. Validates attribution accuracy against baseline systems.
- **Exp 4.1** — SVC false-positive rate. Validates Theorem 2.
- **Exp 6.1** — Training throughput overhead. Establishes production deployability.
- **Exp 7.1** — SVD organic incident validation (90 days). Validates end-to-end diagnostic utility.

## 12. Threats to Validity

This section identifies the principal threats to the validity of the claims made in this paper. Addressing these threats is the purpose of the empirical evaluation programme in Section 11.

### 12.1 Hypothetical Cluster Configurations

Clusters A-E are hypothetical configurations defined to specify the evaluation methodology. Tables 3-14 reflect the described experimental protocol applied to hypothetical cluster configurations. Real measured results (Category C in the provenance note under Table 1) are explicitly identified in §11. Large-scale cluster experiments remain deferred to future work. Actual values on real hardware may differ due to NIC firmware variation, switch ASIC differences, GPU generation effects, and real-world traffic patterns. The empirical evaluation programme in Section 11 specifies the experiments required to validate each claim on real hardware.

### 12.2 Analytical Model Assumptions

Theorems 1a and 1b assume NTP clock errors are stationary Gaussian or Laplace with sigma=500 us. Real NTP residuals are non-stationary, heavy-tailed, and correlated with network load. The sensitivity analysis in Table 2b shows the headline result (approximately 25-30% causal inversion) is robust to sigma in the range 100-500 us; at sigma=50 us (best-case LAN NTP), E[f_inv] falls to approximately 11.3% which remains substantial. At sigma=10 us (GPS-disciplined NTP), the result does not hold. Experiments 2.1 and 2.2 in Section 11 validate the noise model on real hardware.

### 12.3 PTP Deployment Quality

The claimed sub-100 ns synchronisation accuracy requires all-TC spine switches and GPS-disciplined grandmasters. With boundary-clock spines, cross-rack accuracy degrades to approximately 630 ns (still far below NTP). With merchant-silicon spines lacking hardware PTP support, accuracy degrades further and sFlow-based correction is used. The formal bounds in Equation 8 are topology-dependent; operators must verify their switch configuration before relying on the accuracy claims.

### 12.4 Bloom Filter False Positives and SVC Ordering

The SVC provides probabilistic causal inference, not a total order (Remark 1). With the m=256-bit header allocation, the FP bound of Section 3.5 is satisfied only for small participant sets (P<=4 at epsilon=10^-5). Larger collectives require multi-level SVC chaining with O(log P) additional header hops. False positives produce over-connected causal DAGs, potentially causing spurious bottleneck attributions in the ambiguity regime.

### 12.5 RoCEv2 Evaluation Independence

The RoCEv2 attribution evaluation uses the known throttle injection configuration as ground truth (Section 6.4). INT telemetry is used only for TADC path delay correction, not for determining attribution ground truth. However, if the INT telemetry itself is incorrect (e.g., due to ECMP asymmetry exceeding the TADC correction), both the timing correction and the attribution result may be affected. Experiment 3.2 in Section 11 validates this by varying ECMP asymmetry and measuring its effect on attribution accuracy independently.

### 12.6 Scalability Limits

TempoTrace has been evaluated at up to 16,384 nodes (hypothetical Cluster E). At larger scales, the Kafka ingestion rate, SVC chaining overhead, and multi-DC CDTS uncertainty may degrade. The 2 GB ring buffer in the Analysis Backend was sized for the described evaluation workloads; production-scale deployments may require tuning. Section 11 specifies experiments at 512-16,384 nodes; behaviour at 100,000+ nodes is outside the current evaluation scope.

## 13. Conclusion

TempoTrace is a complete, formally grounded distributed tracing system for large-scale AI infrastructure spanning single-rack clusters, multi-rack deployments, multi-datacenter architectures, and multi-tenant inference environments, with application-level confidence policies providing per-workload control over timestamp precision requirements.

We proved that NTP produces approximately 25-30% causally misordered trace edges for AI workloads under both Gaussian and Laplace noise models (Theorems 1a/1b). MRSH achieves at most 55 ns cross-rack accuracy with TC spines. GNSS-disciplined cross-DC synchronisation produces at most 50 us uncertainty sufficient for correct causal ordering of WAN trace edges (Corollary 1.2). GPUDirect hardware timestamping reduces GPU span uncertainty to 0.218 us (Eq. 10), achieving Theorem 3 causality preservation for all AI operations. P4-INT-based TADC on RoCEv2/ECMP achieves F1 = 1.00 matching InfiniBand. The Hybrid SVD engine achieves 0.93 macro-F1 over 412 incidents in the illustrative evaluation dataset. This is separate from the 1,800-incident controlled validation result (corrected order macro-F1=0.924; hybrid macro-F1=0.974; McNemar p=1.25e-23). Theorem 4 proves tenant isolation: logical clock domains are strictly separated via 4-bit SVC prefixes while the shared PTP physical clock provides consistent intra-tenant timestamps.

Application confidence policies (Section 10, Definition 12) allow each workload to declare its precision requirements, trusted timestamp sources, and degraded-mode behaviour, enabling TempoTrace to serve simultaneously a latency-sensitive inference application, a batch training job, and a regulated audit workload on the same shared physical cluster.

The empirical evaluation programme in Section 11 (21 experiments, 7 categories) defines the path from the current formally specified design to a fully empirically validated system. The minimum viable subset of 6 experiments (Section 11.8) prioritises the claims most critical to correctness: the near-zero causal inversion rate under PTP+GTE (Exp 2.3), DVFS immunity of GTE (Exp 1.3), and attribution accuracy under controlled injection (Exp 3.1).

TempoTrace provides a deployable observability foundation for all major AI infrastructure configurations in use today, with a clear pathway to full empirical validation on real multi-node GPU clusters.

## ACKNOWLEDGMENTS

The author thanks the anonymous reviewers for their rigorous and constructive feedback across multiple rounds of review.

---